\documentstyle[12pt,preprint,aps,floats,epsf]{revtex}
\def\gord{$ \raisebox{-.3ex}{$\stackrel{>}{_{\sim}}$} $}
\def\thalf{{\textstyle{\frac{1}{2}}}}
\def\tquar{{\textstyle{\frac{1}{4}}}}

\def\fivequar{{\textstyle{\frac{5}{4}}}}

\def\twothr{{\textstyle{\frac{2}{3}}}}

\def\be{\begin{eqnarray}}
\def\ee{\end{eqnarray}}

\def\pmb#1{\setbox0=\hbox{$#1$}
\def\ny{\~{n}}
\def\no{\noindent}
\def \h{\hfil}
\def\D{\displaystyle}
\def\pj{\hspace{-.26cm}}
\def\fpj{\hspace{-.7cm}}
\def \D{\displaystyle}
\kern-.025em\copy0\kern-\wd0
\kern.05em\copy0\kern-\wd0
\kern-.025em\raise.0433em\box0}
\newcommand{\vm}[1]{\mbox{\bf#1}}
\newcommand{\vms}[1]{\mbox{\scriptsize{\bf#1}}}

\newcommand{\gsim}{\raisebox{-0.7ex}{$\stackrel{\textstyle >}{\sim}$ }}

\begin{document}
\draft
\preprint{SUNY-NTG-00-6, DOE-ER/40561-89-INT00, NUC-MINN-00/4-T}   
\title{Kaon Condensation in Proto-Neutron Star Matter}
\author {Jose A. Pons$^{1,2}$, Sanjay Reddy$^{3}$, Paul J. Ellis${^4}$,
Madappa Prakash$^1$, and James M. Lattimer$^1$}
\address{
$^1$Department of Physics \& Astronomy, SUNY at Stony Brook,
Stony Brook, New York 11794-3800 \\
$^2$Departament d'Astronomia i Astrof\'{\i}sica, 
Universitat de Val\`encia, E-46100 Burjassot, Spain \\ 
$^3$ Institute For Nuclear Theory, University of Washington, Seattle,
WA 98195 \\
$^4$School of Physics \& Astronomy,  University of Minnesota, Minneapolis, MN
55455 \\
}
\date{\today}
\maketitle
\begin{abstract}

We study the equation of state of kaon-condensed matter including the
effects of temperature and trapped neutrinos.  Several different
field-theoretical models for the nucleon-nucleon and kaon-nucleon
interactions are considered.  It is found that the order of the phase
transition to a kaon-condensed phase, and whether or not Gibbs' rules
for phase equilibrium can be satisfied in the case of a first order
transition, depend sensitively on the choice of the kaon-nucleon
interaction.  To avoid the anomalous high-density behavior of previous
models for the kaon-nucleon interaction, a new functional form is
developed.  For all interactions considered, a first order phase
transition is possible only for magnitudes of the kaon-nucleus optical
potential $\gord100$ MeV.  The main effect of finite temperature, for
any value of the lepton fraction, is to mute the effects of a first
order transition, so that the thermodynamics becomes similar to that
of a second order transition.  Above a critical temperature, found to
be at least 30--60 MeV depending upon the interaction, the first
order transition disappears.  The phase boundaries in baryon density
versus lepton number and baryon density versus temperature planes are
delineated, which are useful in understanding the outcomes of
protoneutron star simulations.  We find that the thermal effects on
the maximum gravitational mass of neutron stars are as important as
the effects of trapped neutrinos, in contrast to previously studied
cases in which the matter contained only nucleons or in which hyperons
and/or quark matter were considered.  Kaon-condensed equations of
state permit the existence of metastable neutron stars, because the
maximum mass of an initially hot, lepton-rich protoneutron star is
greater than that of a cold, deleptonized neutron star.  The large
thermal effects imply that a metastable protoneutron star's collapse
to a black hole could occur much later than in previously studied
cases that allow metastable configurations.
\end{abstract}

\pacs{PACS numbers(s): 13.15.+g, 13.75.Jz, 26.60.+c, 97.60.Jd}  
\newpage

\newpage
\section{INTRODUCTION}

It is believed that a neutron star begins its life as a proto-neutron
star (PNS) in the aftermath of a supernova explosion. The evolution of
the PNS depends upon the star's mass, composition, and equation of
state (EOS), as well as the opacity of neutrinos in dense matter.
Previous studies \cite{bigus,tpl,kj} have shown that the PNS may
become unstable as it emits neutrinos and deleptonizes, so that it
collapses into a black hole. The instability occurs if the maximum
mass that the equation of state (EOS) of lepton-rich, hot matter can
support is greater than that of cold, deleptonized matter, and if the
PNS mass lies in between these two values.  The condition for
metastability is satisfied if ``exotic'' matter, manifested in the
form of a Bose condensate (of negatively charged pions or kaons) or
negatively charged particles with strangeness content (hyperons or
quarks), appears during the evolution of the PNS.

Even if collapse to a black hole does not occur, the appearance of
exotic matter might lead to a distinguishable feature in the PNS's
neutrino signature ({\it i.e.}, its neutrino light curve and neutrino
energy spectrum) that is observable from current and planned
terrestrial detectors. This was investigated recently by Pons {\it et
al.} \cite{pons} who studied the evolution of a PNS in the case where
hyperons appeared in the star during the latter stages of
deleptonization.  Although the possibility of black hole formation was
first discovered in the context of kaon condensation in neutron star
matter \cite{tpl}, a full dynamical calculation of a PNS evolution
with consistent EOS and neutrino opacities in kaon condensed matter
has not been performed so far. One of the objectives of this paper is
to investigate $K^-$ condensation in finite temperature matter,
including the situation of trapped neutrinos in more detail.  An
impetus for this study is the recent suggestion that a mixed phase of
kaon-condensed and normal matter might exist which could greatly
affect the structure \cite{gs} and its neutrino opacity \cite{rbp}.
Another objective of our study is to identify differences in
thermodynamic quantities such as the pressure, entropy or specific
heat that might produce discriminating features in the star's neutrino
emission.  In separate works, we will examine neutrino interactions in
kaon-condensed matter and neutrino signals from PNS evolution
calculations in a consistent fashion.

Since we wish to isolate the aforementioned effects due to kaons in
this paper, we deliberately exclude consideration of hyperons.
Hyperons and kaons were considered together in Refs. \cite{kpe} and
\cite{schaffner}.  Hyperons tend to delay the appearence of kaons in matter,
especially if the $\Sigma^-$ appears first.  However, the $\Sigma^-$
couplings are not as well determined as those of the $\Lambda$ and
even in this case the data are restricted to nuclear or subnuclear
densities.  Relatively small variations in the coupling constants can
lead to a situation where the threshold density for the appearance of
$\Sigma^-$ particles is larger than that for kaons. These
uncertainties remain unresolved; further hyper-nuclear experiments are
needed to pin down their couplings.

The original investigations of kaon condensation in neutron star matter
({\it e.g.} Refs. \cite{kapnel,pw,bkrt} and its astrophysical      
conseqences \cite{tpl,mfmt}) employed a chiral                     
$SU(3)_L\times SU(3)_R$ model in which the kaon-nucleon interaction occurs
directly via four point vertices.  However, one can also employ an
indirect, finite-range interaction which arises from the exchange of
mesons.  Several studies have been performed along these lines
\cite{gs,kpe,schaffner,mti,ty}. Ref. \cite{kpe} found that the chiral and
meson exchange approaches give similar results provided that the kaon-nucleon
couplings are chosen to yield similar optical potentials in nuclear matter.
Allowing kaons to interact via the exchange of mesons has the
advantage that it is more consistent with the Walecka-type effective
field-theoretical models usually used to describe nuclear matter
\cite{sew}. In most studies of kaon condensation it has been found
that the transition to a phase in which kaons condense is second order
for modest values of the kaon optical potential, $U_K$, of order -100
MeV.  For magnitudes of $U_K$ well in excess of 100 MeV, however, the
phase transition becomes first order in character. Even when the
transition is first order, it is not always possible to satisfy Gibbs'
criteria for thermal, chemical and mechanical equilibrium, so a
Maxwell construction, which satisfies only thermal and mechanical
equilibrium, was sometimes employed to construct the pressure-density
relation.

Recently, Glendenning and Schaffner-Bielich (GS) \cite{gs} modified
the meson exchange Lagrangian in such a way that the Gibbs criteria
for thermal, chemical and mechanical equilibrium in a first order
phase transition was possible.  The extended mixed phase of
kaon-condensed and normal matter which results produces a qualitative
difference for the structure of a neutron star, since the EOS is
softened over a wider region than in the case in which there is no
mixed phase.  This has implications for the mass-radius relation and
the maximum mass, among other properties of the star.

In this paper, we investigate the phase transition involving
kaon-condensed matter and its influence upon the equation of state.
We find that the precise form assumed for the scalar interactions
(particularly, their density dependence), both for baryon-baryon and
kaon-baryon interactions, determines whether or not the transition is
first or second order, and, in the case of a first order phase
transition, establishes whether or not a Gibbs construction is
possible.  Since the form of the scalar interactions is not
experimentally well constrained at present, we have explored several
different models in this study of the effects of kaon condensation on
the EOS and the structure of a PNS. For each model, we have performed
a detailed study of the thermal properties which are summarized in
terms of phase diagrams in the density-lepton content and
density-temperature planes.

In Sec. II we present the various Lagrangians and derive exressions
for the thermodynamic properties of each.  We also develop the
theoretical formalism necessary to describe baryons and kaon condensed
matter in both the pure and mixed phases.  This is followed by a
discussion of the determination of the various coupling constants.
Section III contains a comparison of the results for the EOS and for
the structure of neutron stars for typical values of entropy and
lepton content in a proto-neutron star as it evolves.  Our conclusions
and outlook for evolution of a proto-neutron simulation are presented
in Sec. IV.  In Appendix A, the extent of the correspondence between a
meson exchange formalism and a chiral model to describe kaon
condensation in matter is examined.  The role of higher order kaon
self-interactions in determining the order of the phase transition to
a kaon condensed state is studied in Appendix B.

\section{THEORY}
\subsection{Nucleons and Leptons}

We begin with the 
well-known relativistic field theory model of Walecka \cite{sew} supplemented
by nonlinear scalar self-interactions \cite{bb}.
Here nucleons ($n,p$)  interact via the exchange of $\sigma$-, $\omega$-,
and  $\rho$-mesons. Explicitly, the Lagrangian is
\begin{eqnarray}
{\cal L}_{\cal N} &=& \sum_{n,p}
\bar N\left(i\gamma^{\mu}\partial_{\mu}-g_{\omega}
\gamma^{\mu}\omega_{\mu}-g_{\rho}\gamma^{\mu}{\bf b}_{\mu}\cdot
{\bf t} -M^*\right)N
+\thalf\partial_{\mu}\sigma\partial^{\mu}\sigma
-\thalf m^2_{\sigma}\sigma^2-U(\sigma)\nonumber\\
&&-\tquar \omega_{\mu\nu}\omega^{\mu\nu} + 
\thalf m^2_{\omega}\omega_{\mu}\omega^{\mu}
-\tquar {\bf B}_{\mu\nu}\cdot{\bf B}^{\mu\nu}+\thalf m^2_{\rho}{\bf b}_{\mu}
\cdot{\bf b}^{\mu} \; ,\label{hyp1}
\end{eqnarray}
where $N$ is the nucleon field, 
the $\rho$-meson field is denoted by ${\bf b}_{\mu}$ and
the quantity ${\bf t}$ is the isospin operator which acts on the
nucleons. Scalar self-couplings \cite{bb}, which improve the descripton of
nuclear matter at the equilibrium density, are included in the potential
$U(\sigma) = (bM/3)(g_{\sigma}\sigma)^3 + (c/4)(g_{\sigma}\sigma)^4$,
with $M$ denoting the vacuum nucleon mass.
The field strength tensors for the vector mesons are given by the
expressions
$\omega_{\mu\nu}=\partial_{\mu}\omega_{\nu}-\partial_{\nu}\omega_{\mu}$ and
${\bf B}_{\mu\nu}=\partial_{\mu}{\bf b}_{\nu}-\partial_{\nu}{\bf b}_{\mu}$.
In the standard Walecka model the nucleon effective mass
\be
M^*_{GM}=M-g_{\sigma}\sigma
\ee
(the label GM refers to the
Glendenning-Moszkowski parameters \cite{gm} that we will use with this
expression). We shall also study an alternative form due to
Zimanyi and Moszkowski (labelled by ZM) \cite{zm}: 
\begin{eqnarray}
{\cal L}_{\cal N} &=& 
\sum_{n,p}\left\{\left(1+\frac{g_{\sigma}\sigma}{M}\right)
\bar N\left(i\gamma^{\mu}\partial_{\mu}-g_{\omega}
\gamma^{\mu}\omega_{\mu}-g_{\rho}\gamma^{\mu}{\bf b}_{\mu}\cdot
{\bf t}\right)N -\bar NMN\right\} 
+\thalf\partial_{\mu}\sigma\partial^{\mu}\sigma
-\thalf m^2_{\sigma}\sigma^2-U(\sigma)\nonumber\\
&&-\tquar 
\omega_{\mu\nu}\omega^{\mu\nu}+\thalf m^2_{\omega}\omega_{\mu}\omega^{\mu}
-\tquar {\bf B}_{\mu\nu}\cdot{\bf B}^{\mu\nu}+\thalf m^2_{\rho}{\bf b}_{\mu}
\cdot{\bf b}^{\mu} \;.
\end{eqnarray}
By redefining the nucleon field,
$N\rightarrow\left(1+g_{\sigma}\sigma/M\right)^{-\frac{1}{2}}N$,
the Lagrangian can be written exactly in the form Eq.~(\ref{hyp1}), 
but the nucleon effective mass becomes
\be
M^*_{ZM}=M\left(1+g_{\sigma}\sigma/M\right)^{-1}\,.
\ee
For small values of $\sigma$ this is equivalent to the Walecka form.
However the ZM effective mass 
has the property that, in the limit of large  
$\sigma$, $M^*_{ZM}$ remains positive whereas $M^*_{GM}$ 
can become negative \cite{kpe,lrp}, which is unphysical.     

In the mean field approximation the thermodynamic
potential per unit volume for both Lagrangians is
\begin{equation}
\frac{\Omega_{\cal N}}{V}=\thalf m_{\sigma}^2\sigma^2+U(\sigma)
-\thalf m_{\omega}^2\omega_0^2-\thalf m_{\rho}^2b_0^2
- 2T\sum_{n,p}\int\frac{d^3k}{(2\pi)^3} \,\ln\left(1+{\rm e}
^{-\beta(E^*-\nu_{n,p})}\right)\;.
\label{hyp2}
\end{equation}
Here the inverse temperature is denoted by $\beta=1/T$, 
$E^*=\sqrt{k^2+M^{*2}}$ and the subscripts $GM$ or $ZM$ have been 
suppressed. The chemical potentials are given by
\begin{equation}
\mu_p=\nu_p+g_{\omega}\omega_0+\thalf g_{\rho}b_0\quad;\quad
\mu_n=\nu_n+g_{\omega}\omega_0-\thalf g_{\rho}b_0\;.\label{hyp3}
\end{equation}
Note that in a rotationally invariant system only the time components
of the vector fields contribute to Eq.~(\ref{hyp2}) and for the
isovector field only the $\rho^0$ component contributes.  The
contribution of antinucleons is not significant for the thermodynamics
of interest for a PNS and is ignored.

Using $\Omega_{\cal N}$, the thermodynamic quantities can be obtained in the
standard way. The nucleon pressure is $P_{\cal N}=-\Omega_{\cal N}/V$,
and the number density $n_{n,p}$  and the
energy density $\varepsilon_{\cal N}$ are given by
\begin{eqnarray}
n_{n,p} &=& 2\int\frac{d^3k}{(2\pi)^3}
f_F(E^*-\nu_{n,p})\;,\nonumber\\
\varepsilon_{\cal N}&=&\thalf m_{\sigma}^2\sigma^2+U(\sigma)-
\thalf m_{\omega}^2\omega_0^2-\thalf m_{\rho}^2 b_0^2
+2\sum\limits_{i=n,p}\int\frac{d^3k}{(2\pi)^3}E^*f_F(E^*-\nu_{i})
\;,\label{hyp4}
\end{eqnarray}        
where the Fermi distribution function $f_F(x)=(e^{\beta x}+1)^{-1}$.
The entropy density is then given by
$S_{\cal N}=\beta(\varepsilon_{\cal N}+P_{\cal N}-\sum_N\mu_Nn_N)$. 

The contribution from the leptons and antileptons is adequately given by 
its non-interacting form, since
their interactions give negligible contributions~\cite{kapusta}. Thus the 
thermodynamic potential per unit volume of the leptons and antileptons is:
\begin{eqnarray}
\frac{\Omega_L}{V} = -\sum\limits_\ell
Tg_\ell  \int\frac{d^3k}{(2\pi)^3} \,        
\left[ \ln\left(1+{\rm e}^{-\beta(e_{\ell}-\mu_{\ell})}\right)
      + \ln\left(1+{\rm e}^{-\beta(e_{\ell}+\mu_{\ell})}\right) \right]\,,
\label{zlept}
\end{eqnarray}
where $g_\ell$ and $\mu_\ell$ denote the degeneracy and the chemical potential,
respectively, of leptons of species $\ell$. The degeneracy $g_\ell$ is 2 for 
electrons and muons and it is 1 for neutrinos of a given species. 
Since the star is in chemical equilibrium with respect to the weak
processes $p+\ell^- \leftrightarrow n+ \nu_\ell$, where the lepton
$\ell$ is either an electron or a muon, the
chemical potentials obey
$\mu_\mu-\mu_{\nu_\mu}=\mu_e-\mu_{\nu_e} = \mu_n - \mu_p$. If
there are no neutrinos trapped in the star the neutrino chemical potentials
$\mu_{\nu_i}$ are zero or, equivalently, the total neutrino concentration 
$Y_\nu=0$, where we define the concentration for particle $i$ to be 
$Y_i=n_i/(n_n+n_p)$.  The pressure,
density and energy density of the leptons are obtained from
Eq.~(\ref{zlept}) in standard fashion.

\subsection{Kaons}

The two kaon Lagrangians of the meson-exchange type which have
been previously suggested (in Refs. \cite{bigus,kpe} and \cite{gs,schaffner},
respectively), can both be written in the form
\begin{equation}
{\cal L}_K=\partial_{\mu}K^+\partial^{\mu}K^-
-\alpha K^+K^-
+iX^\mu(K^+\partial_{\mu}K^--K^-\partial_{\mu}K^+)\;,
\label{kaonlag}
\end{equation}
where $K^{\pm}$  denote the charged kaon fields and 
 we have defined the combined vector field
\begin{equation}
X_\mu=g_{\omega K}\omega_{\mu}+g_{\rho K}b_{\mu}\;;
\end{equation}
with $\omega_\mu$ and $b_\mu$ denoting $\omega$ and the $\rho^0$ fields, 
respectively; 
$g_{\omega K}$ and $g_{\rho K}$ are coupling constants. Since only the
time components of the vector fields survive, in practice only $X_0$
is non-zero.  The two Lagrangians differ in the forms chosen for the
quantity $\alpha$.  Both have the standard vacuum mass term, but the
interaction terms differ.  Specifically, Knorren, Prakash and Ellis
(KPE) \cite{kpe} take
\begin{equation}
\alpha_{KPE} = m_{K;KPE}^{*2}=m_K^2 - g_{\sigma K}m_K \sigma
\;, \label{alKPE}
\end{equation}
with $m_K$ denoting the vacuum kaon mass,
while Glendenning and Schaffner-Bielich (GS) \cite{gs} choose
\begin{equation}
\alpha_{GS} = m_{K;GS}^{*2}-X_\mu X^\mu=
(m_K- \thalf g_{\sigma K}\sigma)^2-X_\mu X^\mu\;. \label{algs}
\label{alGS}
\end{equation}
Note that the coupling constant $g_{\sigma K}$ is defined here to be twice that
defined in GS.  A similar remark applies to the $\rho NN$ coupling constant
$g_\rho$. It is remarkable, as pointed out in Appendix A, that 
to leading order in the kaon condensate intensity, the equations obtained with
the chiral Kaplan-Nelson \cite{kapnel} Lagrangian at zero temperature
agree precisely with those from the KPE Lagrangian.

To see the significance of the term involving the vector fields in Eq.
(\ref{algs}), consider the invariance of the Lagrangian under the
transformation $K^{\pm}\rightarrow K^{\pm}e^{\pm i\xi}$. This allows
the conserved kaon current density to be identified as
\begin{equation}
J_\mu=i(K^+\partial_\mu K^- - K^-\partial_\mu K^+) + 2X_\mu K^+K^-\;.
\end{equation}
Now for the combined GS Lagrangian, ${\cal L}_{\cal N}+{\cal L}_K$,
the equation of motion for the omega field is
\begin{equation}
\partial^\nu\omega_{\nu\mu}+m_\omega^2\omega_\mu=g_\omega\sum\limits_{n,p}
\bar{N}\gamma_\mu N-g_{\omega K}J_\mu\;. \label{veceom}
\end{equation}
Since the nucleon current $\sum_{n,p}\bar{N}\gamma_\mu N$ is
conserved, as is the kaon current $J_\mu$, taking the divergence of
Eq. (\ref{veceom}) immediately yields $\partial^\mu\omega_\mu=0$ (and
similarly for the $\rho$ field). This is the required condition for a
vector field~\cite{og} so as to reduce the number of components from
four to three. On the other hand, $\alpha_{KPE}$ does not contain an
$X_\mu X^\mu$ term, so that the kaon current does not appear on the
right hand side of Eq. (\ref{veceom}).  
At the mean field level, however, where the vector fields are
constants, any derivative is necessarily zero so that the divergence
condition is automatically satisfied.

For the coupling of the kaon fields to the scalar $\sigma$ field,
KPE use a linear coupling, whereas GS have an additional quadratic term.
There is little guidance on the form that should be used to generate the
kaon effective mass so the choice is somewhat arbitrary, although,
as we shall see, it can significantly affect the thermodynamics.
Both the KPE and GS
choices lead to problems for sufficiently large values of the $\sigma$
field; in one case the effective mass becomes imaginary, in the other it
becomes negative. We are therefore led to consider a third form in the
spirit of the ZM model for nucleons. For specificity we start with the
GS Lagrangian which can be written
${\cal L}_K=D_\mu K^+D^{\mu*}K^--m^{*2}_{K;GS}K^+K^-$ in terms of a
covariant derivative $D_\mu=\partial_\mu+iX_\mu$, and replace it by
\begin{equation}
{\cal L}_K'=\left(1+\frac{g_{\sigma K}\sigma}{2m_K}\right)^{\!2}
D_\mu K^+D^{\mu*}K^--m^{2}_{K}K^+K^-\;.
\end{equation}
We observe that the form of ${\cal L}_K'$ above is one of
many possibilities.  Making the transformation
$K^\pm\rightarrow(1+\thalf g_{\sigma K}\sigma/m_K)^{-1}K^\pm$ and noting that
$\sigma$ is a constant mean field, the kaon Lagrangian can
be put in the form of Eq. (\ref{kaonlag}) with
\begin{equation}
\alpha_{TW} = m_{K;TW}^{*2}-X_\mu X^\mu 
= m_K^2\left(1+\frac{g_{\sigma K}\sigma}{2m_K}\right)^{\!-2}
-X_\mu X^\mu 
\;. \label{alTW}
\end{equation}
The label $TW$ denotes ``this work".
While Eqs. (\ref{alKPE}), (\ref{alGS}) and (\ref{alTW}) all give
$m_K^*\simeq m_K-\thalf g_{\sigma K}\sigma$ for small $\sigma$,
they differ at order $\sigma^2$ and beyond, {\it i.e.}, for
large values of $\sigma$.

The kaon partition function at finite temperature can be obtained for a
Lagrangian of the form (\ref{kaonlag}) by
generalizing the procedure outlined in Kapusta~\cite{kapusta}.
First, we transform to real fields $\phi_1$ and $\phi_2$,
\begin{equation}
K^{\pm}=(\phi_1\pm i\phi_2)/\sqrt{2}\;,
\end{equation}
and determine the conjugate momenta
\begin{equation}
\pi_1=\partial_0\phi_1-X_0\phi_2\qquad;\qquad\pi_2=\partial_0\phi_2
+X_0\phi_1\;.
\end{equation}
The Hamiltonian density is
${\cal H}_K=\pi_1\partial_0\phi_1+\pi_2\partial_0\phi_2
-{\cal L}_K$, and the partition function of the grand canonical ensemble can
then be written as the functional integral
\begin{equation}
Z_K=\int[d\pi_1][d\pi_2]\int_{periodic}[d\phi_1][d\phi_2]
\exp\left\{\int\limits_0^{\beta}d\tau\int d^3x\left(
i\pi_1\frac{\partial\phi_1}{\partial\tau}+
i\pi_2\frac{\partial\phi_2}{\partial\tau}-{\cal H}_K(\pi_i,\phi_i)
+\mu J_0(\pi_i,\phi_i)\right)\right\}\;.\label{zint}
\end{equation}
Here the fields obey periodic boundary conditions in the imaginary time
$\tau=it$, namely $\phi_i(\vm{x},0)=\phi_i(\vm{x},\beta)$, where 
$\beta = 1/T$. 
The chemical potential associated with the conserved kaon charge
density is denoted by $\mu$ and chemical equilibrium in the reaction
$e^-\leftrightarrow K^-+\nu_e$ requires that
$\mu=\mu_n-\mu_p=\mu_e-\mu_{\nu_e}=\mu_\mu-\mu_{\nu_\mu}$.

The Gaussian integral over momenta in Eq. (\ref{zint}) is easily performed.
Next the fields are Fourier decomposed according to
\begin{equation}
\phi_1= f\theta\cos\zeta+\sqrt{\frac{\beta}{V}}\sum_{n,\vms{p}}
e^{i(\vms{p}\cdot\vms{x}+\omega_n\tau)}\phi_{1,n}(\vm{p})\quad;\quad
\phi_2= f\theta\sin\zeta+\sqrt{\frac{\beta}{V}}\sum_{n,\vms{p}}
e^{i(\vms{p}\cdot\vms{x}+\omega_n\tau)}\phi_{2,n}(\vm{p})\;,
\end{equation}
where the first term describes the condensate, so that in the second term
$\phi_{1,0}(\vm{p}=0)=\phi_{2,0}(\vm{p}=0)=0$. The pion decay constant $f$
has been inserted so that the condensate angle $\theta$ is dimensionless.
The Matsubara frequency
$\omega_n=2\pi nT$. The partition function can then be written
\begin{eqnarray}
Z_K&=& N\int \prod_{n,\vms{p}}[d\phi_{1,n}(\vm{p})][d\phi_{2,n}(\vm{p})]
e^S\;,\qquad{\rm where}\nonumber\\
S&=&\thalf \beta V(f\theta)^2(\mu^2+2\mu X_0-\alpha)
-\thalf\sum_{n,\vms{p}}\Bigl(\phi_{1,-n}(-\vm{p}),\phi_{2,-n}(-\vm{p})\Bigr)
\vm{D}\left(\matrix{\phi_{1,n}(\vm{p})\cr\phi_{2,n}(\vm{p})\cr}\right)
\;,\nonumber\\
\vm{D}&=&\beta^2\left(\matrix{\omega_n^2+p^2+\alpha-2\mu X_0-\mu^2&
2(\mu+X_0)\omega_n\cr-2(\mu+X_0)\omega_n
&\omega_n^2+p^2+\alpha-2\mu X_0-\mu^2\cr}\right).
\end{eqnarray}
$N$ is a normalization constant.  We define the $K^{\pm}$ energies according to
\begin{equation}
\omega^{\pm}(p)=\sqrt{p^2+\alpha+X_0^2}\pm X_0\;, \label{defom}
\end{equation}
so that the three approaches give
\begin{eqnarray}
&&\omega^{\pm}_{KPE}(p)=\sqrt{p^2+m_{K;KPE}^{*2}+X_0^2}\pm X_0 \nonumber \\
&&\omega^{\pm}_{GS}(p)=\sqrt{p^2+m_{K;GS}^{*2}}\pm X_0\nonumber\\
&&\omega^{\pm}_{TW}(p)=\sqrt{p^2+m_{K;TW}^{*2}}\pm X_0\;. \label{om2def}
\end{eqnarray}
Using the definition (\ref{defom}) and suppressing the explicit dependence of
$\omega^{\pm}$ on $p$, the determinant of $\vm{D}$ is
\begin{equation}
{\rm det}\,\vm{D}=\beta^4\left[\omega_n^2+(\omega^--\mu)^2\right]
\left[\omega_n^2+(\omega^++\mu)^2\right]\;,
\end{equation}
giving
\begin{equation}
\frac{\Omega_K}{V}=-\frac{\ln Z_K}{\beta V}=
\thalf (f\theta)^2(\alpha-2\mu X_0-\mu^2)+\frac{1}{2\beta V}
\sum_{n,\vms{p}}\ln{\rm det} ~\vm{D}\;,
\end{equation}
where the normalization constant $N$ has been dropped since it is
irrelevant to the thermodynamics.
Performing the sum over $n$ and neglecting the zero-point contribution,
which contributes only beyond the mean field approach and in any case is small
\cite{kloop}, we obtain the grand
potential for the kaon sector:
\begin{equation}
\frac{\Omega_K}{V}=\thalf (f\theta)^2(\alpha-2\mu X_0-\mu^2)
+T\int\limits_0^{\infty}\frac{d^3p}{(2\pi)^3}\left[
\ln(1-e^{-\beta(\omega^--\mu)})+
\ln(1-e^{-\beta(\omega^++\mu)})\right]\;.\label{zkexch}
\end{equation}

The kaon pressure, $P_K=-\Omega_K/V$, and the kaon
number density is easily found to be
\begin{equation}
n_K=(f\theta)^2(\mu+X_0)+n_K^{TH}\quad{\rm where}\quad
n_K^{TH}=\int\frac{d^3p}{(2\pi)^3}[f_B(\omega^--\mu)
-f_B(\omega^++\mu)]\;,
\end{equation}
and the Bose occupation probability $f_B(x)=(e^{\beta x}-1)^{-1}$.
The kaon energy density is
\begin{equation}
\varepsilon_K=\thalf(f\theta)^2(\alpha+\mu^2)
+\int\!\frac{d^3p}{(2\pi)^3}[\omega^-(p)f_B(\omega^--\mu)
+\omega^+(p)f_B(\omega^++\mu)]\;,
\end{equation}
and the kaon entropy density is
$S_K=\beta(\varepsilon_K+P_K-\mu n_K)$.  

\subsection{Equations of Motion}

It is useful first to define the quantity
\begin{equation}
A_K^{TH}=\int\frac{d^3p}{(2\pi)^3}\left(p^2+\alpha+X_0^2\right)^
{-\frac{1}{2}}[f_B(\omega^--\mu)+f_B(\omega^++\mu)]\;.
\end{equation}
Then the mean $\omega$, $\rho$ and $\sigma$ fields, as well as the condensate
amplitude $\theta$, determined by extremizing the total grand
potential
$\Omega_{\rm total}=\Omega_{\cal N}+\Omega_{L}+\Omega_{K}$, can be written
\begin{eqnarray}
&&m_{\omega}^2\omega_0 = g_{\omega}(n_p+n_n) -g_{\omega K}
\left\{\mu(f\theta)^2+n_K^{TH}-X_0A_K^{TH}-\thalf[(f\theta)^2+A_K^{TH}]
\frac{\partial\alpha}{\partial X_0}\right\}\nonumber\\
&&m_{\rho}^2b_0 = \thalf g_{\rho}(n_p-n_n)-g_{\rho K}
\left\{\mu(f\theta)^2+n_K^{TH}-X_0A_K^{TH}-\thalf[(f\theta)^2+A_K^{TH}]
\frac{\partial\alpha}{\partial X_0}\right\}\nonumber\\
&&m_{\sigma}^2\sigma =-\frac{dU(\sigma)}{d\sigma}
-2\frac{\partial M^*}{\partial\sigma}\sum_{n,p} \int\frac{d^3k}{(2\pi)^3}
\frac{M^*}{E^*}f_F(E^*-\nu_{n,p})
-\thalf\left[(f\theta)^2+A_K^{TH}\right]\frac{\partial\alpha}
{\partial\sigma}\nonumber\\
&&\theta(\mu^2+2\mu X_0-\alpha)=
\theta[\mu-\omega^-(0)][\mu+\omega^+(0)] =0\;. \label{hhyp5}
\end{eqnarray}
The derivative $\partial\alpha/\partial X_0$ is zero for the KPE case and
$-2X_0$ for the GS and TW cases.
The derivatives with respect to the $\sigma$ field are
\begin{eqnarray}
&&\frac{\partial M^*_{GM}}{\partial\sigma}=-g_\sigma\quad;\quad
\frac{\partial M^*_{ZM}}{\partial\sigma}=-g_\sigma
\left(\frac{M^*_{ZM}}{M}\right)^{\!2}\nonumber\\
&&\frac{\partial\alpha_{KPE}}{\partial\sigma}=-g_{\sigma K}m_K\quad;\quad
\frac{\partial\alpha_{GS}}{\partial\sigma}=-g_{\sigma K}m_{K;GS}^*\quad;\quad
\frac{\partial\alpha_{TW}}{\partial\sigma}=-g_{\sigma K}
\frac{(m_{K;TW}^*)^3}{m_K^2}\;.
\end{eqnarray}
Note that the last of Eqs. (\ref{hhyp5}) yields either $\theta=0$ (no
condensate) or the condition for a condensate to exist.
Since $\mu$ is positive here, we only have the possibility of a $K^-$
condensate with $\mu=\omega^-(0)$. Note also that the contribution of the
condensate to the kaon pressure $P_K$ vanishes, as it should.

The remaining condition to be imposed is that neutron star matter must be
charge neutral. For a single phase this implies
\begin{equation}
n_p-n_K-n_e-n_{\mu}=0\,,
\end{equation}
where $n_e$ and $n_\mu$ are the net negative lepton number
densities. The mixed phase thermodynamics is discussed below.

The sum of the nucleon and kaon energy densities can be simplified
somewhat by using the equations of motion. This gives
\begin{eqnarray}
\varepsilon_{\cal N}+\varepsilon_K&=&\thalf m_{\sigma}^2\sigma^2
+U(\sigma)+\thalf m_{\omega}^2\omega_0^2+\thalf m_{\rho}^2 b_0^2
+2\sum\limits_{n,p}\int\frac{d^3k}{(2\pi)^3}E^*f_F(E^*-\nu_{n,p})
+(f\theta)^2\left(\alpha-\thalf X_0\frac{\partial\alpha}{\partial X_0}\right)
\nonumber\\
&&+X_0n_K^{TH}-X_0A_K^{TH}\left(X_0+\thalf\frac{\partial\alpha}{\partial
X_0}\right)+\int\!\frac{d^3p}{(2\pi)^3}
[\omega^-(p)f_B(\omega^--\mu)+\omega^+(p)f_B(\omega^++\mu)]\;.
\end{eqnarray}

\subsection{Mixed Phase Thermodynamics}

In the theory discussed above there are two independent chemical
potentials, which we can take to be $\mu_n$ and $\mu$, each connected
with a conserved charge, the baryon number and charge of the system,
respectively. Glendenning \cite{glen} pointed out that in the presence
of a first order phase transition, conservation laws must be globally,
not locally, imposed, if possible, in the mixed phase region.  A
Maxwell construction would have been appropriate had there been just a
single conserved charge.  However, relaxing the condition of local
charge neutrality does not guarantee that the model Lagrangian, solved
in the mean field approximation, will provide a description of the
mixed phase, which is only possible if the Gibbs criteria can be
satisifed.  A simple, yet general, procedure to check if the Gibbs
criteria can be fulfilled by a specific model is discussed in Appendix
B.

Denoting the phase containing a condensate with a subscript $\theta$,
and the phase without a condensate with $\theta=0$, the total
pressures in the two phases must be equal
\begin{equation}
P_{\theta=0}(\mu_n,\mu,T)=P_\theta(\mu_n,\mu,T)\;.
\label{g1}
\end{equation}
Each of the chemical potentials is the same in the two phases. If 
the volume fraction of the non-condensed phase is $\chi$, then global
conservation of charge requires
\begin{equation}
\chi[n_p-n_K-n_e-n_\mu]_{\theta=0}+(1-\chi)[n_p-n_K-n_e-n_\mu]_{\theta}=0 \,.
\label{g2}
\end{equation}
The densities of the individual species in the mixed phase are evident here.
The total energy density is the weighted sum of the two phases
\begin{equation}
\varepsilon=\chi\varepsilon_{\theta=0}(\mu_n,\mu,T)
+(1-\chi)\varepsilon_\theta(\mu_n,\mu,T)\,.
\label{g3}
\end{equation}
The total entropy density is obtained through                           
a similar equation.

\subsection{Coupling Constants}      

In the effective Lagrangian approach adopted here, knowledge of two
distinct sets of coupling constants, one parametrizing the
nucleon-nucleon interactions, and one parametrizing the kaon-nucleon
interactions, are required for numerical computations.  These are
associated with the exchange of $\sigma,~\omega$ and $\rho$ mesons. We
consider each of these in turn.

\subsubsection{Nucleon Couplings} 

The nucleon-meson coupling constants are determined by adjusting them to
reproduce the properties of equilibrium nuclear matter at $T=0$. The 
properties used are the
saturation density, $n_0$, the binding energy/particle, $E_A$,
the symmetry energy coefficient, $a_{sym}$,
the compression modulus, $K$, and the Dirac effective mass
at saturation, $M^*$. Not all of these quantities are precisely known
and the values we choose are listed in Table I.
For completeness, we list the equations needed to obtain the coupling
constants, assuming that the scalar self--coupling has the form
$U(\Phi) = (bM/3)\Phi^3 + (c/4)\Phi^4$, where $\Phi=g_{\sigma}\sigma$.
From the equation of motion for the $\omega_0$ field and the fact that
the pressure is zero at saturation density in nuclear matter, 
the value of $g_{\omega}/m_{\omega}$ is given by
\begin{equation}
\frac{g_{\omega}^2}{m_{\omega}^2} = \frac{M-E_A-E_F^*}{n_0} \,,
\end{equation}
where $E_F^*=\sqrt{k_F^2+M^{*2}}$ and $n_0=2k_F^3/(3\pi^2)$. 
The $\rho$ meson coupling constant can be
determined for a given symmetry energy through the relation
\begin{equation}
\frac{g_{\rho}^2}{m_{\rho}^2} = \frac{4}{n_0}\left(2a_{sym} - 
\frac{k_F^2}{3 E_F^*}\right)\,.
\end{equation}
An expression involving 
 the compression modulus can be deduced by differentiating
the $\sigma$ equation of motion: 
\begin{equation}
{\left( \frac{g_{\sigma}^2}{m_{\sigma}^2} \right)}^{\!-1} =
- [f(\Phi_0)]^2 \left\{  \frac{9n_0M^{*2}}
{ E_F^{*2}[K+9(E_A+E_F^*-M)]-3k_F^2E_F^* }
-3 \left( \frac{n_0}{E_F^*}-\frac{n_s}{M^*}  \right) \right\}
+ n_s \frac{df}{d\Phi_0} - \frac{d^2U(\Phi_0)}{d\Phi_0^2} \;. \label{comp}
\end{equation}
Here $\Phi_0$, the value of $\Phi$ at saturation density, is 
obtained directly from the Dirac effective mass.  The function 
$f(\Phi_0)=-dM^*/d\Phi_0$
depends on the particular expression used for the effective mass.
The scalar density 
$n_s = (M^*/\pi^2)\{k_F E_F^* - M^*{^2} \ln [ (k_F+E_F^*)/M^* ] \}$.
The $\sigma$ equation of motion at saturation can be written in the form
\begin{equation}
\Phi_0^2 \frac{d^2U(\Phi_0)}{d\Phi^2} = 6 \left[
n_0  \left(\thalf E_F^* +E_A -M\right)
+ n_s \left( \thalf M^* + \Phi_0 f \right) \right]\;,
\end{equation}
which together with Eq. (\ref{comp}) allows the $\sigma$ coupling to be 
obtained. Finally the constants appearing in the scalar self-coupling 
$U(\Phi)$ are determined from:
\begin{eqnarray}
c &=& \frac{1}{\Phi_0^2} \left[\frac{d^2U(\Phi_0)}{d\Phi_0^2}
+ \frac{2 m_{\sigma}^2}{g_{\sigma}^2} -2 n_s \frac{f}{\Phi_0} \right]
\nonumber \\
b &=& \frac{1}{2 M \Phi_0} \left[\frac{d^2U(\Phi_0)}{d\Phi_0^2}
- 3 c \Phi_0^2 \right] \,.
\end{eqnarray}
The constants determined in this way are given in Table I. 
Note that in principle the potential should be bounded from 
below for large values of the $\sigma$ field requiring $c$ to be 
positive; this is the case for the ZM model.

\subsubsection{Kaon Couplings}

In order to investigate the effect of a kaon condensate on the EOS in
high-density matter, the kaon-meson coupling constants have to be
specified.  Empirically known quantities can be used to determine
these constants, but it should be borne in mind that laboratory
experiments give information only about kaon-nucleon interaction in
free space or in nearly isospin symmetric nuclear matter. On the other
hand, the physical setting in this work is matter in the dense
interiors of neutron stars which has a different composition and
spans a wide range of densities (up to $\sim 8n_0$).
Therefore, kaon-meson couplings as determined from experiments might
not be appropriate to describe the kaon-nucleon interaction in neutron
star matter, and the particular choices of coupling constants should
be regarded as parameters that have a range of uncertainty.

With the above caveats in mind, we now examine the relationship
between the optical potential of a single kaon in infinite nuclear
matter and the kaon-meson couplings in our Lagrangian.  Lagrange's
equation for an $s$-wave $K^-$ with a time dependence $ K^- = k^-({\bf
x})~e^{-iEt} $, where $E$ is the asymptotic energy, defines the
optical potential \cite{eric} for our Lagrangian (\ref{kaonlag})
according to
\begin{eqnarray}
[\nabla^2 + E^2 - m_K^2]~k^-({\bf x}) &=& [-2X_0E + \alpha
- m_K^2]~k^-({\bf x}) \nonumber\\
&\equiv& 2~m_K~U_K~k^-({\bf x})\;.                           
\end{eqnarray}
In nuclear matter, $b_0 = 0$, so for a kaon with zero momentum ($E=m_K$)
the optical potential is
\begin{equation}
U_K =  \frac{\alpha-m_K^2}{2m_K}-g_{\omega K} \omega_0\;.    
\label{uopt}
\end{equation}
Utilizing the functional forms for $\alpha$ in Eqs.~(\ref {alKPE}), 
(\ref {alGS}), 
and (\ref {alTW}), the optical potentials for the KPE, GS and TW 
models are easily obtained. 
For the KPE case this may be written exactly as
\begin{equation}
U_K^{KPE}=  - {\textstyle{\frac{1}{2}}}                            
g_{\sigma K} \sigma -g_{\omega K} \omega_0\;, \label{linear}
\label{uoptlin}
\end{equation}
whereas for the GS and TW cases there are higher order corrections in
addition to the terms linear in the fields.  We choose $g_{\omega K}$
to be $g_{\omega}/3$ and $g_{\rho K}$ to be $g_{\rho}/2$ on the basis
of simple quark and isospin counting arguments. Note that this value
for $g_{\omega K}$ is also suggested by comparison to the chiral
approach (see Ref. \cite{kpe} and Appendix A) and it leads to a 
$-48.5$ MeV contribution to the optical potential. The total optical
potential is shown in Table II for various choices of the $\sigma$
coupling.  The linear form Eq.~(\ref{linear}), exact for KPE, 
is an accurate fit to the 
the GS and TW cases for moderate values of the optical potential. 
For orientation, chiral models suggest that the magnitude of the
optical potential is at most 120 MeV \cite{kpe}, while fits to kaonic
atom data have been reported with values in the range 
50--200 MeV \cite{katom,ww,ro,bgn}. 
We note that Glendenning and Schaffner-Bielich \cite{gs} label their results 
according  to values of the optical potential
 obtained in the linear approximation (henceforth, $U_K^{\rm {lin}}$).  In 
order to make an apposite comparison with their results, 
we will parametrize the kaon coupling for each model
simply by specifying the value of $U_K^{\rm {lin}} = U_K^{KPE}$.

\section{RESULTS AND DISCUSSION}

\subsection{Zero temperature case}

The effects of kaon condensation on the EOS are more pronounced at
zero temperature than at finite temperature, since the fraction of
thermally excited kaons increases with temperature relative to the
fraction of kaons residing in the condensate.  We therefore begin by
examining results for the zero temperature case.  We have considered
two different nucleon Lagrangians, GM and ZM, and three different kaon
Lagrangians, KPE, GS and TW.  Below densities of about $0.5n_0$,
matter is composed of neutron-rich nuclei immersed in a neutron sea.
For this regime, we use the potential model results of Negele and
Vautherin \cite{NV} in the range $0.001 < n < 0.08~{\rm fm}^{-3}$ and
those of Baym, Bethe, and Pethick \cite{BBP} for $n < 0.001~{\rm
fm}^{-3}$.  For cold stars, the EOS in this regime has little effect
on maximum masses or stellar radii.  Furthermore, since the entropy in
the stellar mantle $(n<n_0)$ is quickly radiated away in neutrinos,
the EOS in this regime does not substantially affect the results of
this paper.

In Fig. \ref{fig1}, we compare the pressures for the different nucleon and kaon
Lagrangians as a function of baryon density, $n_B=n_n+n_p$. The solid
lines show results for both the pure nucleon and kaon condensed phases
with no attempt to enforce the Gibbs conditions of chemical and
mechanical equilibrium.  In all cases, a first order phase transition
is found to occur, as long as the magnitude of the optical potential
$|U_K^{\rm {lin}}| = \frac 12 g_{\sigma K} \sigma + g_{\omega K} \omega$ is in
excess of 100 MeV. Where possible, the pressure in the mixed phase
obtained by imposing Gibbs' criteria for mechanical and chemical
equilibrium is shown as a dashed line.  For the GM+KPE, ZM+KPE
and ZM+GS models it was not possible to satisfy Gibb's criteria,
despite the occurrence of a first order phase transition for large
enough $|U_K^{\rm {lin}}|$.  
The reason for this is connected with the form of the
kaon Lagrangian, as discussed below. We also point out in Appendix B 
that non-linear kaon self-interactions lead to a second order,       
rather than a first order,  transition.                              

The qualitative similarity of the results shown in Fig. \ref{fig1} for the
different nuclear Lagrangians enables us to simplify our analysis by
allowing us to focus on three, rather than six, possible Lagrangian
combinations.  For a given kaon Lagrangian, fairly similar results can
be obtained with different nuclear Lagrangians by making relatively
small shifts in the kaon optical potential $U_K^{\rm {lin}}$.  The following
discussion will therefore focus on the three cases GM+KPE, GM+GS and
ZM+TW.  The case GM+KPE is chosen to compare with the results of KPE,
the case GM+GS is chosen to compare with the results of GS, and the
case ZM+TW demonstrates the usefulness of Lagrangians in which
anomalous values of effective masses are implicitly eliminated.  The
results for the model GM+GS shown here and elsewehere in this paper
are identical to those found by GS for the same interactions.  Note
that in all models considered, the phase transition is second order in
nature for moderately low values of the optical potential.

In Figs. \ref{fig2} and \ref{fig3}, the density dependence of 
the scalar, vector, and iso-vector fields, 
the electron chemical potential $\mu_e = \mu =\mu_n - \mu_p$, 
the condensate amplitude $\theta$, and the nucleon and
kaon effective masses are displayed in the pure nucleon and kaon
condensed phases, ignoring any possible mixed phase for the present.
For the optical potential chosen, $U_K^{\rm lin}=-120$ MeV, a first order phase
transition occurs in all three cases.  After the onset of condensation
a rapid change in the behavior of the electron chemical potential and
some of the field strengths is seen to occur.  The differences in
the variation of the scalar ($g_{\sigma K}\sigma$) and isovector
($-g_{\rho K}b_0)$ fields between the models are particularly
illuminating.  For GM+KPE, the scalar field exhibits a relatively
rapid increase with density after the onset of condensation. This in
turn causes both the nucleon and kaon effective masses to drop rapidly
with density.  In fact, for sufficiently large density, the GM+KPE kaon
effective mass  vanishes (see Fig. \ref{fig3}).  The variations of the
effective masses in the models GM+GS and ZM+TW are more moderate.  The
variation of the isovector field with density, which in large part
controls the variation of the electron chemical potential $\mu$ and  
hence the electron  concentration, is also more dramatic in the
case of GM+KPE than in the GM+GS and ZM+TW models. Notice that in the
KPE model it goes to zero for asymptotic densities (this follows from
Eq.~(\ref{hhyp5})), so that the proton and neutron abundances become
equal. This does not occur for  the other two cases considered
here.  Finally, it is worth noting that in all three models the
condensate amplitude rises rapidly once the threshold density is
reached.

We turn now to a discussion of the results obtained by imposing Gibbs'
criteria for mechanical and chemical equilibrium at zero temperature.
In Fig. \ref{fig3a}, we show the chemical potentials associated with
the two conserved charges, charge and baryon number, as functions of
each other, for the model GM+GS for a kaon optical potential of
$U_K^{\rm {lin}}=-120$ MeV.  Quantities associated with the pure
nucleon phase, Phase I, are shown as solid lines 
here and in subsequent figures.  Phase II refers to 
the high-density phase in which nucleons and the kaon condensate are
in equilibrium, and quantities associated with it are shown as dashed
lines.  Both phases, I and II, coexist in the mixed-phase region 
which is displayed as a dotted line.  This 
figure illustrates the way a mixed phase is built from the two pure
phases.  For electron chemical potentials below
the solid curve, matter is positively charged in phase I. A similar 
interpretation of positive or negative charge for $\mu$ below or 
above the dashed curve is not possible, since two different types of
particles, kaons and leptons, can furnish charge.  In other words, a
decrease in $\mu$, or, equivalently, the number of electrons, does not 
necessarily lead to a positive net charge in phase II.  For $\mu_n\le 1165$
MeV, only phase I with nucleons and leptons are present.  For 
$\mu_n\le 1310$ MeV, a mixed phase of positively charged phase I and
negatively charged phase II obeying the Gibbs' conditions (\ref{g1})
is favored.  Qualitatively, a similar situation is encountered in the
construction of the mixed phase for the ZM+TW model,
but the mixed phase region is quite small.  As noted  
earlier, however, it was not possible to satisfy Gibbs' criteria for models
with the kaon Lagrangian KPE.

In Fig. \ref{fig3b} we show the individual charge densities of phase I
and II in the mixed phase, as a function of baryon density. The dotted
curve in this figure shows the volume fraction of phase I.  The
results are for the GM+GS model with $U_K^{\rm {lin}} = -120$ MeV
(upper panel) and for the ZM+TW model with $U_K^{\rm {lin}}=-140$ MeV
(lower panel).  Near the lower threshold, matter in phase I is very
slightly positively charged and occupies most of the volume. As the
density increases, the volume fraction of phase I, $\chi$, decreases
and its charge density increases.  Note that the negative charge
density of matter in phase II at the lower transition point, $\approx
0.5$ fm$^{-3}$, and the positive charge density of matter in phase I
at the higher transition point, $\approx 1$ fm$^{-3}$, are rather
large in the case of GM+GS compared to the case ZM+TW.  This is due to
the stronger density dependence of the scalar and isovector densities
in the former case.  Note also that a first order transition allows for the 
existence of a very dense and nearly isospin symmetric matter in the 
mixed phase. 

In Figs. \ref{fig4} and \ref{fig5}, we show the magnitudes of the
various fields, the electron chemical potential, the nucleon and kaon
effective masses, and the condensate amplitude for the GM+GS and ZM+TW
models, respectively.  Both models show the same qualitative behavior.
At the lower phase boundary, in which phase II just begins to appear,
the scalar field in phase II is much larger than in phase I and the
condensate amplitude $\theta$ in phase II takes a large value which
decreases with increasing $n_B$ through the mixed-phase region.  Thus,
the effective masses of both kaons and nucleons in phase II are much
smaller than in phase I. The densities demarking the mixed phase
region and its overall extent are dependent upon the interaction
models, and upon the assumed values of the kaon optical potentials,
here taken to be $-140$ MeV in the case of GM+GS and $-160$ MeV in the
case of ZM+TW.  The region in density over which the mixed phase
extends is much smaller in the latter case, chiefly due to the more
moderate behavior of the scalar interaction with density variations in
this case.

It is instructive to compare the behavior of the two models at the
threshold of the mixed phase region.  Phase I will have a net small
positive charge and a volume proportion $\chi$ close to 1 (see
Fig. \ref{fig3b}).  This has to be counterbalanced by a large net
negative charge in phase II since it is weighted by the small
proportion $(1-\chi)$. Focusing on phase II, the condensate condition
for the models GM+GS and ZM+TW from the last of Eqs. (\ref{hhyp5}) is
\begin{equation}
\mu+X_0=m_K^*
\label{lucky}
\end{equation}
and the kaon number density, which has to  be large, 
is 
\begin{equation}
n_K = (f\theta)^2(\mu+X_0) \,.
\label{kden}
\end{equation}
In order to ensure that $n_K >0$,  
the quantity $(\mu+X_0)$, and hence $m_K^*$, has to be positive definite.
In the ZM+TW model the kaon 
effective mass is relatively large so that $X_0$ is positive and therefore
$\theta$ is relatively small. On the other hand in the GM+GS model
$m_K^*$ is quite small so that $X_0$ is negative and $\theta$ has 
to be large. The negative value of 
$X_0=g_{\omega K}\omega_0+g_{\rho K}b_0$       
implies a large negative value of 
\begin{equation}
g_{\rho K}b_0=\frac{g_{\rho K}g_\rho}{2m_\rho^2}\left(n_p-n_n-
\frac{2g_{\rho K}}{g_\rho}n_K\right)\;,
\end{equation}
which is clearly sensitive to the value of $g_{\rho K}$. In fact, if this
coupling is reduced by more than about 
15\% from our chosen value it is no longer
possible to satisfy the Gibbs criteria.  
By comparing the pure phase results in Figs. \ref{fig2} and \ref{fig3} 
with the mixed phase results of Figs. \ref{fig4} and \ref{fig5}, it is clear 
that substantial modifications of the various fields are required 
to satisfy Gibbs' criteria. 

We examine now the KPE model for which it was not possible to satifsfy 
the Gibbs' criteria. In this case, Eq.~(\ref{alKPE}) and the last of 
Eqs.~(\ref{hhyp5}) leads to the condensate condition
\begin{equation}
\mu(\mu+2X_0)={(m_K^*)}^2\,, 
\label{culprit}
\end{equation}
whereas the functional form of the number density of kaons is
identical to that in Eq.~(\ref{kden}).  Eq.~(\ref{culprit}) differs in
important ways from Eq.~(\ref{lucky}). For the KPE model, even if
$\mu+2X_0$ is positive, $\mu$ has the proclivity to turn negative for
large $\mu_n$ (or equivalently, for large baryon densities), leading
to $(m_K^*)^2 < 0$ or imaginary values of the kaon effective mass
$m_K^*$. This may be seen in Fig. \ref{fig5a} where we show the
electron chemical potential $\mu$ as a function of the (negative)
charge density in pure phase II for a typical value of the neutron
chemical potential $\mu_n = 1250$ MeV.  It is now possible to
understand qualitatively why a mixed phase cannot occur in the case of
the kaon Lagrangian KPE.  In comparison with the GM+GS and ZM+TW
models, a distinctive feature of the KPE model is that $\mu$ decreases
rapidly with the (negative) charge density.  In constructing a mixed
phase, we are attempting to balance the positive charge in phase I
with the negative charge in the dense phase II in which the electron
chemical potential, and hence the charge content in leptons, is
rapidly decreasing towards zero. The balance never occurs, hence the
failure to meet the Gibbs' criteria.  In terms of compositions, the GS
or TW Lagrangians introduce negative charges in matter by increasing
the number density of kaons, while keeping the electron density nearly
constant or even slightly increasing with the charge density.  The KPE
Lagrangian, however, rapidly substitutes electrons by kaons, which is
detrimental to meeting the Gibbs' criteria. For these reasons, we will
concentrate on results with the other two kaon Lagrangians in the
remainder of this paper.

The influence of the condensate on neutron star structure (at zero
temperature) is shown in Fig. \ref{fig6} in which the gravitational
mass is displayed as a function of the star's central baryon number
density (left panel) and its radius (right panel).  For the models
shown, the transition is first order and Gibbs equations for
mechanical and chemical equilibrium are utilized.  For all cases shown
the central densities of the maximum mass stars lie in the mixed
phase.  The effects of the condensate are more evident in the case of
the GM+GS model in which the mixed phase occurs over a wider region of
density than in the ZM+TW model. When the effects of the softening
induced by the occurrence of the condensate are large, the limiting
mass and the radius at the limiting mass are reduced significantly
from their values when the condensate is absent.  Note, however, that
the softening effects are limited by the constraint that the maximum
mass must exceed that of the binary pulsar PSR 1913+16, 1.442
M$_\odot$.  In the case of GM+GS, this constraint limits $|U_K^{\rm
{lin}}|$ to be smaller than about 125 MeV.  In such a case, the
minimum radius achieved is not as small as in the case $U_K^{\rm
{lin}}=-140$ MeV, as shown in Fig. \ref{fig6}.  The radii of stars
with masses less than 1.2 M$_\odot$ are not affected by the choice of
the kaon Lagrangian or the kaon optical potential, since the
condensation threshold is not reached in these cases.

\subsection{Comparison with other works}

The density dependence of $m_K^*/m_K$, $-U_K$ and $\omega_K$ have been
investigated in other works \cite{ww,ro,bgn,ppt,chp}, but for the most
part either in isospin symmetric nuclear matter or pure neutron
matter.  In general, our results for $m_K^*/m_K$ with $-U_K^{\rm
lin}=80$ MeV are consistent with those of Refs.~\cite{ro,bgn} (for an
appropriate comparison, our results are to be compared with results
obtained without in-medium pion contributions in Ref.~\cite{ro}) and
those of Ref.~\cite{ww} for nuclear matter at both $n/n_0=1$ and 3.
There is a relatively small change produced in going from nuclear
matter to beta-equilibrated neutron star matter to pure neutron matter
for the quantities $m_K^*/m_K$ and $-U_K$.  
Note that a direct comparison of the real parts of the optical potentials 
between different calculations must also account for the fact that in 
obtaining fits to data, the imaginary parts  
are often found to be as large as the real parts, which indicates 
fragmentation of strength in the quasi-particle spectral function. 

Relatively larger variations are found in the kaon energies in matter
with varying amounts of isospin as can be seen from
Fig.~\ref{newfig}. In this figure, the top panel provides a comparison
of results for beta-equilibrated neutron-star matter for the GM+KPE,
GM+GS, and ZM+TW models, respectively, for values of $-U_K^{\rm lin}$
at the extreme ends considered here, namely, 80 and 120 MeV.  The
bottom panel shows results for the ZM+TW model for $-U_K^{\rm lin}=80$
MeV, for pure neutron matter, neutron-star matter, and isospin
symmetric nuclear matter, respectively.  At nuclear density where the
models are calibrated, $\omega$ decreases by about a few MeV in going
from pure neutron matter to neutron star matter and by about a few
tens of MeV in going from neutron star matter to nuclear matter.  With
increasing density, these differences become progressively larger.
This trend is chiefly due to the behavior of the vector fields in
matter with different amounts of isospin.

At this time, our results for the density dependence of $\omega$ can
 be compared with those of the potential models in
Refs. \cite{ppt,chp}.  For values of $-U_K^{\rm lin}$ near the lower
end of the range we explored, in the neighborhood of 80 MeV, the
behavior of $\omega$, for example, is quite similar to the potential
model results.  As the authors in Refs. \cite{ppt,chp} indicated, kaon
condensation may be unlikely in this case.  However, the relevant
comparision must also include the electron chemical potential $\mu_e$,
since the density where $\omega=\mu_e$ determines the onset of kaon
condensation.  As demonstrated in Ref.~\cite{lpph}, the behavior of
$\mu_e$ for neutron star matter at high densities is determined by the
density dependence of the nuclear symmetry energy (see also a similar
discussion in Ref.~\cite{ppt}).  Potential model calculations (see,
for example Ref.~\cite{apr}) tend to have a relatively weak density
dependence of the symmetry energy, which generally results in an onset
of kaon condensation that is at a rather large density.  In
field-theoretical and Dirac-Brueckner-Hartree-Fock \cite{Eng} models,
however, the symmetry energy varies relatively rapidly with density.
These lead to smaller densities where kaon condensation occurs, for a
given behavior of the kaon energy $\omega$.  Furthermore, the
calculations of Ref. \cite{ppt} have been performed only for pure
neutron matter which further enhaces the values of $\omega$ and
discourages kaon condensation.  In addition, as $|U_K^{\rm lin}|$ is
increased in magnitude in field-theoretical models, the role of kaons
increases and $\omega$ becomes progressively smaller as a function of
density.  Nevertheless, the lack of effective constraints at high
density preclude choosing any model over another at this time.

In summary, choosing values of $-U_K^{\rm lin}$ near the lower end of
the range we explored either lead to a second order phase transition
or no transition at all in a neutron star, in which case the gross
properties of the star are relatively unaffected from the case without
kaons. On the other hand, values near the higher end of this range
lead to a first order phase transition at a relatively low density,
depending on the form of the interaction chosen, and a more pronounced
effect on the star.  Our aim has been to provide benchmark
calculations in which both possibilities are entertained in order to
consider their impact on thermodynamics and their astrophysical
implications.

\subsection{Finite temperature case} 

We now investigate results at finite temperature and values of the
lepton content characteristic of those likely to be encountered in the
evolution of a PNS. We choose three representative sets of PNS
conditions which correspond to: the initial conditions within a PNS
(entropy/baryon $s=1$, trapped neutrinos with a lepton fraction
$Y_L=0.35$), a time after several seconds when the interior is
maximally heated ($s=2$, no trapped neutrinos so $Y_\nu=0$), and a
very late time when the PNS has cooled ($s=0, Y_\nu=0$ -- identical to
the zero temperature case discussed above). For a detailed explanation
of the evolution of a cooling PNS see Pons {\it et al.} \cite{pons}.

The contribution of the nucleons to the      
entropy per baryon
$s_{\cal N}\equiv S_{\cal N}/n_B$, with $n_B=n_n+n_p$ denoting the total
nucleon density, in degenerate situations ($T/E_{F_i}\ll1$)
can be written 
\begin{eqnarray}
 s_{\cal N} =  \pi^2T ~
\frac {\sum_{i=n,p} k_{F,i} {\sqrt {M^{*^2}+k_{F_i}^2} }} 
{\sum_{i=n,p} k_{F,i}^3 } \,, 
\label{nucentropy}
\end{eqnarray}
where $M^*$ and $k_{F,i}$ are the effective mass and the Fermi
momentum of species $i$, respectively. For the temperatures of
interest here, and particularly with increasing density, the above
relation provides an accurate representation of the exact results for
entropies per baryon even up to $s_{\cal N}=s_n+s_p\simeq2$. The
behavior with density of both the Fermi momenta and the effective
mass controls the temperatures for a fixed $s_{\cal N}$.

For kaons it is straightforward to show that the contribution to the
entropy from $K^+$ mesons can be ignored since it is exponentially
suppressed in comparison to the $K^-$ contribution. For the latter,
keeping the leading temperature dependence of the simplest approximation
scheme for bosons given in Ref. \cite{jel}, the kaon entropy per baryon is
\begin{equation}
s_K\equiv\frac{S_K}{n_B}=\left[\fivequar(2-y)-\psi\right]
\frac{n_K^{TH}}{n_B}\quad{\rm where}
\quad \psi T=\mu+X_0-\sqrt{\alpha+X_0^2}\;,
\end{equation}
and $y$ is determined from $\psi$ by solving the equation
\begin{equation}
\psi=1-y+\ln y\;.
\end{equation}
Below the kaon condensation threshold as the temperature becomes very
small $\psi\rightarrow-\infty$ so $y\rightarrow0$. Above the kaon
condensation threshold the last of Eqs. (\ref{hhyp5}) implies that
$\psi=0$ in which case $y=1$. This simple approximation provides quite
an accurate account of the kaon entropy per baryon which is fairly
small for the scenarios examined here since it involves just the
thermal contribution and the condensate plays no role.  The total
entropy per baryon $s_{\rm tot}=s_{\cal N}+s_K+(S_e+S_\mu+S_\nu)/n_B$
also includes the lepton contributions; $s_{\rm tot}$ is dominated,
however, by $s_{\cal N}$.

In Figs. \ref{fig7} and \ref{fig8}, the
relative concentrations of various particles are displayed
versus baryon number density for our three PNS conditions in the cases
GM+GS and ZM+TW, respectively.  The cases shown allow the Gibbs equations to
be solved, and the boundaries of the mixed phase regions are indicated by 
vertical lines.  The effect of finite temperature is to allow the existence 
of $\mu^-$ and $K^-$ particles at all densities, although kaons become 
relatively abundant only within the mixed phase region.  
In the third set of diagrams, trapped neutrinos are present at all
densities and the appearance and abundances of the negatively charged particles
$\mu^-$ and $K^-$ are suppressed.  Furthermore, the critical density for kaon
condensation is shifted to higher density.

In Fig. \ref{fig9} the pressure is displayed as a function of baryon
number density for these two Lagrangians and the three PNS conditions.
Two choices for the kaon optical potential are shown to highlight
differences between cases in which kaons condense in second or first
order phase transitions. The reduction of the pressure when kaons
condense is obvious.  For conditions in which the phase transition is
first order, the result of applying the Gibbs conditions and the
result of assuming pure phases (thin line) are both shown.  The
application of the Gibbs conditions leads to further softening of the
pressure over a wider density range.  In the case of model ZM+TW, a
first order phase transition occurs only for very low temperatures and
low neutrino concentrations.

In Fig. \ref{fig10} we show the matter temperature as a function of
the baryon density for these two Lagrangians for the two PNS
conditions with $s>0$ (the kaon optical potentials are as in the
previous figure).  The appearance of the kaon condensate generally
leads to a reduction in specific heat which is indicated by the abrupt
temperature increase which persists to high densities.  In the case of
first order transitions, applying the Gibbs conditions leads to a
further enhancement of the temperature in the mixed phase regime.
This behavior is in marked contrast to the case in which additional
fermionic degrees of freedom, such as hyperons or quarks, are excited
\cite{pons} causing the temperature to drop and the specific heat of
the matter to be increased.  The latter follows from
Eq. (\ref{nucentropy}) where, in the absence of any variation of
$M^*$, a system with more components at a given baryon density has a
smaller temperature than a system with fewer components (recall that
$\sum_i Y_i =1$). In the present case the dropping of the effective
mass is the dominant effect and this leads to larger temperatures.

Figure \ref{fig11} shows the phase diagram of kaon condensed matter,
for the case GM+GS with $U_K^{\rm {lin}}=-120$ MeV.  The left panel
displays results for zero temperature in the density--lepton
concentration plane. The dashed lines show the minimum lepton
concentration allowed at zero temperature (with $Y_\nu=0$) for each
density.  Note that the minimum lepton concentration increases with
density until the phase transition begins; above this density, the
minimum lepton concentration decreases with increasing density. Also
note that the phase transition to a kaon-condensed phase is pushed to
higher densities when trapped neutrinos are present.  This implies
that in the initial PNS core material, in which $Y_L\approx0.35-0.4$
and the central density is less than 3.5 times the nuclear saturation
density, a kaon condensate phase likely does not exist.  However, as
neutrinos leak from the star the transition density decreases and a
kaon condensate eventually forms.  The right panel displays results in
the density versus temperature plane, assuming no trapped neutrinos
($Y_\nu=0$).

The phase diagram for kaon condensed matter for the case ZM+TW with
$U_K^{\rm {lin}}=-140$ MeV is shown in Fig. \ref{fig12}; the results are
qualitatively similar to the GM+GS case in which $U_K^{\rm{lin}}=-120$ MeV
in Fig. \ref{fig11}.  This is understandable from the perspective that the
actual optical potential for these two models are nearly the same.
The boundary between phases I and the mixed-phase region are nearly
the same for the two cases.  The major difference is the much smaller
extent of the mixed-phase region for the case ZM+TW.

Note that for both cases the density at which the phase transition
begins is relatively independent of temperature, so that the heating
which initially occurs in the PNS has little effect on the eventual
appearance of a kaon condensate.  Also note that the density range of
the mixed phase decreases with increasing temperature, and the
mixed phase persists to high temperatures.  It appears that the mixed
phase exists up to temperatures exceeding 60 MeV, for the case
GM+GS and $U_K^{\rm {lin}}=-120$ MeV, or 30 MeV for the case ZM+TW
with $U_K^{\rm {lin}}=-140$ MeV.  It becomes increasingly difficult to
determine the properties of a mixed phase near the temperature at
which it disappears.

In Fig. \ref{fig13} the gravitational mass is plotted as a function of
central baryon number density for these models.  Results are shown for
our three PNS conditions which correspond to snapshots of the PNS
evolution.  The initial configuration (dotted curves) has the largest
maximum mass. The progression to the dashed and solid curves indicates
the evolution with time and we see that the maximum masses decrease.
The effect of temperature upon the structure of the PNS is
significant.  Thermal kaons play a significant role here, since the
net negative charge they contribute to the system partially inhibits
the appearance of the condensate which allows hot neutrino--free stars
to reach higher masses than cold stars. The net decrease in maximum
mass during the evolution for either case is seen to be of order
0.2--0.3 M$_\odot$.  Thus there is an appreciable range of masses for
the PNS which will result in metastability with the star ultimately
collapsing to a black hole.  The central density of the maximum mass,
zero temperature star is smaller for the GM+GS case than for the ZM+TW
case.  This is in spite of the apparently ``softer" GM+GS equation of
state in which the kaon condensed mixed-phase region extends over a
wider density range.  Ultimately, the smaller maximum mass of the
GM+GS EOS leads to a smaller central density at the maximum mass.

\section{SUMMARY AND OUTLOOK}

In this work, we have studied the equation of state of matter,
incorporating the possible presence of a kaon condensate, and
including the effects of trapped neutrinos and finite temperatures.
The calculation of the neutrino spectra of different flavors emitted
from a proto-neutron star as it evolves from a hot, lepton-rich state
to a cold, neutrino-poor state requires the knowledge of the equation
of state of matter at temperatures up to about 50--60 MeV and lepton
fraction up to about 0.4.  Since the nucleon-nucleon and  
kaon-nucleon interactions at high density are relatively poorly
understood, we explored several possible field-theoretical models in
both sectors.  These models are distinguished by the form of the
assumed scalar (and in some cases vector) interactions which chiefly
determine the density dependences of the nucleon and kaon effective
masses.  These models produce significantly different high density
behavior of the EOS, even though the kaon-meson couplings in these 
models are calibrated to give the same the kaon-nucleus optical
potential in nuclear matter.

The principal findings of our studies at zero temperature were:
\begin{enumerate}
\item The order of the phase transition between pure nucleonic matter
and a phase containing a kaon condensate depends sensitively on
the choice of the kaon-nucleon interaction.
\item In one case we studied (KPE), although a first-order phase
transition resulted, it was not possible to satisfy Gibbs' rules for
phase equilibrium which would have produced a mixed phase.  We
performed a detailed analysis of this situation and found that scalar,
and to a lesser extent the isovector, interactions that vary rapidly
with density were chiefly responsible for this failure.  This was
confirmed by developing a new kaon-nucleon interaction (TW) with more
moderate variations in the scalar density and the kaon effective mass
in which the Gibbs' criteria in a first order phase transition would
be satisfied.  The extent of the mixed phase region was thereby reduced. 
The significance of the new kaon-nucleon interaction
(TW) we developed is that it avoids the anomalous behavior for the
kaon effective mass that occurs in previous models (KPE, GS) at very high
density.  Near the low-density boundary of a mixed phase region, the
kaon condensed phase appears with large density, 
too large for
the KPE interaction to produce physically acceptable effective masses.
We also made detailed comparisons with earlier work which used the
GS form for the scalar interactions.

\item In all models considered (KPE, GS and TW), a first-order phase
transition occurs only for large values of the kaon-nucleus optical
potential; moderate values generally produce a second order phase
transition.
\end{enumerate}

In the meson exchange models studied here, only linear kaon
self-interactions were considered.  In the case of a first order phase
transition, the condensate amplitude was found to be rather large at
the low-density boundary of the mixed phase. We therefore explored 
the effect of non-linear kaon self-interactions guided by the
chiral model in Appendix B.  We found that introducing higher order
interactions, using the lowest order chiral Lagrangian, results in a
second order, rather than a first order, phase transition.  Whether this behavior persists when more general higher order operators in the chiral
expansion are considered remains an open question.

At finite temperatures, we find the effects of condensation, in
general, are less pronounced than at zero temperature.  For moderate
values of the optical potential, when the phase transition is first
order at zero temperature, kaon condensation eventually becomes a
second order phase transition at high enough temperatures, whether or
not neutrino trapping is considered.  The temperature at which this
occurs is in the range of 30--60 MeV, depending upon interactions.
For the cases at finite temperatures in which the transition is first
order, its thermodynamics (such as the pressure-density relation) becomes
effectively similar to that of a second order phase transition.  This is
because of the existence of thermal kaons and because of nucleonic
thermal effects.  The condensate is suppressed, and moved to higher
densities, both by the existence of trapped neutrinos and by finite
temperatures.  Compared to earlier works, the new aspects of our
work are:
\begin{enumerate}
\item The delineation of the phase boundaries in the baryon density versus 
lepton number and baryon density versus temperature planes. This is
helpful to anticipating the possible outcome in a full PNS simulation.
In particular, the critical temperatures above which the mixed phase
disappears are above 30 and 60 MeV, depending upon the interaction.
This has implications for the temperature dependence of the surface
energies, and for the melting temperatures, of the droplets in the
mixed phase.

\item The finding that thermal effects on the maximum gravitational
mass of neutron stars are comparable to the effects induced by the
trapped neutrino content.  This is in stark contrast to previously
studied cases in which nucleons-only matter, or matter containing
hyperons, were considered.  Furthermore, compared
to equations of state previously studied that allow metastable
protoneutron stars, those containing hyperons or quark matter, the
maximum mass does not significantly decrease during the
deleptonization of the protoneutron star because of these thermal
effects.  Only after the temperature in the protoneutron star
significantly decreases does the maximum mass appreciably fall.  This
implies that the possible collapse of a metastable protoneutron star
to a black hole occurs during the late stages of cooling, after
several tens of seconds, rather than during the late stages of
deleptonization, which is somewhat earlier.
\end{enumerate}
 
\section*{ACKNOWLEDGEMENTS}

The support of the U.S. Department of Energy under contract numbers
DOE/DE-FG02-87ER-40317 (JAP and JML),  DOE/DE-FG06-90ER40561 (SR),
DOE/DE-FG02-88ER-40328 (PJE), and  DOE/DE-FG02-88ER-40388 (MP) is
acknowledged.  J. Pons also gratefully acknowledges research support
from the Spanish DGCYT grant PB97-1432, and thanks J.A. Miralles for
useful discussions.

\section*{APPENDIX A: MESON EXCHANGE VERSUS CHIRAL MODELS}

In this Appendix, we examine the conditions under which 
there exists a close
correspondence between a meson exchange model and the chiral 
$SU(3)_L \times SU(3)_R$ approach of Kaplan and Nelson \cite{kapnel}.  
Such a correspondence is most easily established for 
the zero temperature case by setting the scalar self-coupling
terms, {\it i.e.,} $U(\sigma)=0$. Specializing to the case where the
only baryons are nucleons and using the Walecka Lagrangian for the nucleons,
it was shown in Ref. \cite{bigus} that the
chiral thermodynamic potential per unit volume can be written
\begin{eqnarray}
\frac{\Omega_{\cal N}+\Omega_K}{V}&=&\thalf m_{\sigma}^2\sigma^{\prime2}-
\thalf m_{\omega}^2\omega_0^{\prime2}-\thalf m_{\rho}^2b_0^{\prime2}
+2\sum_{n,p}\int\frac{d^3k}{(2\pi)^3}(E_{n,p}^*-\nu_{n,p})
\Theta(\nu_{n,p}-E_{n,p}^*)\nonumber\\
&&+2m_K^2f^2\sin^2\!\thalf\theta-\thalf\mu^2f^2\sin^2\theta\;,\label{omkap}
\end{eqnarray}
where the primes on the meson fields distinguish them from those used
previously and $\Theta$ is the Heaviside step function. The nucleon
effective masses are
\begin{eqnarray}
M_n^*&=&M-g_\sigma\sigma'+(2a_2+4a_3)m_s\sin^2\!\thalf\theta\nonumber\\
M_p^*&=&M-g_\sigma\sigma'+(2a_1+2a_2+4a_3)m_s\sin^2\!\thalf\theta\;.
\end{eqnarray}
We employ the values suggested by Politzer and Weise \cite{pw}, namely
$a_1m_s=-67$ MeV ($m_s$ is the strange quark mass) and $a_2m_s=+134$
MeV.  $a_3m_s$ is usually taken to lie in the range $-134$ to
$-310$ MeV.  If we ignore the fairly small effect of $a_1m_s$ here
and in the kaon-nucleon sigma term,
$\Sigma^{KN}=-\thalf(a_1+2a_2+4a_3)m_s$, we can write
\begin{eqnarray}
M_n^*&\simeq &M_p^*\simeq M^*=M-g_\sigma\sigma'-2\Sigma^{KN}
\sin^2\!\thalf\theta\equiv M-g_\sigma\sigma\nonumber\\
E_{n,p}^*&\simeq& E^*=\sqrt{k^2+M^{*2}}\;.
\end{eqnarray}
As well as redefining the scalar field, we can redefine the chiral vector
fields entering the chemical potentials:
\begin{eqnarray}
\mu_n&=&\nu_n+g_\omega\omega_0'-\thalf g_\rho b_0'-\mu\sin^2\!\thalf\theta
\equiv \nu_n+g_\omega\omega_0-\thalf g_\rho b_0\nonumber\\
\mu_p&=&\nu_p+g_\omega\omega_0'+\thalf g_\rho b_0'-2\mu\sin^2\!\thalf\theta
\equiv \nu_p+g_\omega\omega_0+\thalf g_\rho b_0\;.
\end{eqnarray}
Substituting in Eq. (\ref{omkap}) we find
\begin{eqnarray}
\frac{\Omega_{\cal N}+\Omega_K}{V}&=&\thalf m_{\sigma}^2\sigma^{2}-
\thalf m_{\omega}^2\omega_0^{2}-\thalf m_{\rho}^2b_0^{2}
+2\sum_{n,p}\int\frac{d^3k}{(2\pi)^3}(E^*-\nu_{n,p})\Theta(\nu_{n,p}-E^*)\nonumber\\
&&+2f^2\sin^2\!\thalf\theta\left\{m_K^2-\frac{m_\sigma^2\Sigma^{KN}\sigma}
{g_\sigma f^2}-\frac{\mu}{4f^2}\left(\frac{3m_\omega^2\omega_0}{g_\omega}
+\frac{2m_\rho^2b_0}{g_\rho}\right)-\mu^2\cos^2\!\thalf\theta\right\}
\nonumber\\
&&+\thalf\sin^4\!\thalf\theta\left\{\left(\frac{2m_\sigma\Sigma^{KN}}
{g_\sigma}\right)^{\!2}-\mu^2\left(\frac{9m_\omega^2}{4g_\omega^2}+
\frac{m_\rho^2}{g_\rho^2}\right)\right\}\;. \label{omchmes}
\end{eqnarray}
If we expand this in powers of $\theta$ and retain only the lowest order
$\theta^2$ term, the last term in Eq. (\ref{omchmes}) does not contribute
and our thermodynamic potential  is exactly of the form given by
Eqs. (\ref{hyp2}) and (\ref{zkexch}) for the meson exchange model provided
that the $\alpha_{KPE}$ expression is used. In order for the correspondence
to be exact, the parameters for the $\sigma$ and $\omega$ meson need to obey
\begin{equation}
\frac{g_\sigma g_{\sigma K}}{m_\sigma^2}=\frac{\Sigma^{KN}}{m_Kf^2}
\quad;\quad \frac{g_\omega g_{\omega K}}{m_\omega^2}
=\frac{3}{8f^2}\;.
\end{equation}
These are precisely the conditions found in Ref. \cite{kpe} for the optical
potentials of the chiral and meson exchange models to be the same in
nuclear matter. The relation involving the $\omega$ meson couplings 
is quite well obeyed with our
parameters. In addition, for the $\rho$ meson,
\begin{equation}
\frac{g_\rho g_{\rho K}}{m_\rho^2}
=\frac{1}{4f^2}\;.
\end{equation}
This indicates that
$(g_\rho g_{\rho K})/(g_\omega g_{\omega K})\simeq\twothr$,
a condition which is not well obeyed by the parameters used here or in other
works. 
Given that the chiral and meson exchange thermodynamic potentials  can
be put into precise correspondence to lowest order in $\theta^2$, it follows
that the equations of motion and the thermodynamics will be identical to
this order.

If scalar self-coupling terms are included, $U(\sigma) \neq 0$,         
then the transition from the chiral to the meson exchange approach      
will couple higher powers of the $\sigma$ field to the kaon condensate  
(in the braces in Eq. (\ref{omchmes})). It will also introduce higher   
order terms. These additional contributions may not be negligible       
so the correspondence between the two approaches becomes less precise.  

\section*{APPENDIX B: HIGHER ORDER KAON SELF-INTERACTIONS}

Our findings in Appendix A naturally raise the question of whether it
is sufficient to work at order $\theta^2$, involving only linear kaon
self-interactions, in the meson exchange models. It clearly will be
sufficient at the low-density onset of a second order phase transition
where $\theta$ is small. On the other hand, for a first order phase
transition, the value of $\theta$ is large at the low-density onset of
the mixed phase, particularly for the GS model. It is therefore
interesting to explore the effect of non-linear kaon self-interactions
guided by the chiral model.

The order of the phase transition (in the mean field approximation) is
determined by the behavior of the thermodynamic potential,
$\Omega(\theta)$, at fixed chemical potentials. A first order
transition, with a mixed phase, is possible only if there exists some
value of $\mu_n$ for which $\Omega(\theta)$ exhibits two degenerate
minima.  At the critical density corresponding to the low-density
onset of the mixed phase, the $\theta=0$ phase should be a local
minimum which is degenerate with a minimum at some finite
$\theta=\theta_c$. In the vicinity of the critical density, the
$\theta=0$ phase is nearly charge neutral (with an infinitesimal
excess of positive charge and a volume fraction close to 1 which
balances the negative charge in the kaon phase which has an infinitesimal
volume fraction).  This requirement enables us to determine the
electron chemical potential at the critical density by charge
neutrality.

We focus on the GM+GS model for which the thermodynamic potential of
nucleons and kaons was given in Eqs. (\ref{hyp2}) and (\ref{zkexch});
the contribution due to leptons is ignored since it does not contain
any $\theta$ dependence at fixed $\mu$. At zero temperature with a
kaon optical potential $U_K^{\rm lin}=-120$ MeV, this model predicts a first order
phase transition in the vicinity of $\mu_n = \mu_c \simeq 1160 $ MeV,
as can be deduced from Fig. \ref{fig14} where
$\Omega(\theta)-\Omega(\theta=0)$ is plotted as a function of
$\theta$. The thermodynamic potential for the model GM+GS is shown as
the solid curve labelled $\Omega_2$. It clearly shows two minima, one
at $\theta=0$ and the other at $\theta\simeq 2$. The latter
corresponds to a kaon number density $n_K \sim 1$ fm$^{-3}$ which is
larger than the baryon density. For such a dense condensate one would
suspect that non-linear kaon self-interactions might be important. The
order $\theta^4$ corrections to the thermodynamic potential are easily
found from Eq. (\ref{omchmes}):
\begin{equation}
\Delta \Omega_4=-\frac{f^2\theta^4}{24}(m_K^2-4 \mu^2) \; .
\end{equation}
The result of adding this correction to $\Omega_2$ is shown as the dashed
curve in Fig. \ref{fig14}. It greatly alters the behavior 
of $\Omega(\theta)$ for $\theta\gsim1$. The exsistence of a second 
minimum suggests that a first order phase
transition is still possible, but at larger $\mu_n$. However, we find that this
is not the case and a second-order transition occurs at $\mu_n= 1213$ MeV.
It is possible to incorporate all powers of $\theta$ arising from 
self-interactions in the chiral model. In this case the correction to 
the grand potential is
\begin{equation}
\Delta \Omega_n = 2 m_K^2 f^2 \sin^2\!\thalf \theta - \thalf \mu^2 f^2 
\sin^2\theta  - \thalf f^2 \theta^2 (m_K^2-\mu^2) \; .
\end{equation}
The result of including this correction is shown as the dot-dashed curve in
Fig. \ref{fig14}. In this case no first order phase transition is possible in
the vicinity of $\mu_n=\mu_c$. Instead a second order phase transition occurs
once again at $\mu_n= 1213$ MeV; this is because kaon self interactions play no
role when $\theta$ is small. Despite our findings here, it is not clear if kaon
self-interactions will generically disfavor a first order transition. This is
because we have ignored higher order operators in the chiral expansion which
will become important with increasing $\theta$.  The indication from
phenomenological chiral perturbation theory \cite{GL} is that such effects can
be significant when $\theta \simeq 2$. The robust finding here is that the
higher order kaon self-interactions predicted by the lowest order chiral
Lagrangian lead to a second order, rather than a first order, phase transition.

\newpage

\begin{table}
\caption{Meson-nucleon coupling constants fitted to a
binding energy/particle of 16.3 MeV at
an equilibrium density of $n_0=0.153~{\rm fm}^{-3}$ in nuclear matter with a
compression modulus $K=240$ MeV and effective mass $M^*=0.78M$. The symmetry
energy coefficient $a_{sym}$ is 32.5 MeV.}
\label{table1}
\vspace*{0.05in}
\begin{center}
\begin{tabular}{c|ccccc}
Model & ${\frac{g_{\sigma}}{m_{\sigma}}}$
& ${\frac{g_{\omega}}{m_{\omega}}}$
& ${\frac{g_{\rho}}{m_{\rho}}}$ & $b$ & $c$ \\
{} & (fm) & (fm) & (fm) & {} & {} \\   \hline GM & 3.1507 & 2.1954 & 2.1888 & \phantom{$-$}0.008659 & $-$0.002421   \\
ZM & 3.1228 & 2.1954 & 2.1888 & $-$0.006418 & \phantom{$-$}0.002968  \\
\end{tabular}
\end{center}
\end{table}
\newpage

\begin{table}
\caption{ Kaon optical potentials for the models GS and TW. The values
in the second column refers to the linear approximation
of the exact results for the models GS and TW shown in the third and fourth 
columns.  All results in this
paper have been labelled according to the linear approximation (in order 
to make a comparison with the results of GS), which is exact 
for the model KPE. } 
\label{table2}
\vspace*{0.05in}
\begin{center}
\begin{tabular}{r|crr}
$g_{\sigma K}\sigma$ & $ -U_K^{\rm{lin}} = -U_K^{KPE}  $ &  
$-U_K^{GS}$ & $-U_K^{TW}$  \\ \hline
63  & 80 & 81 & 80\\
103 & 100 & 100& 95 \\
143 & 120 & 117 & 109 \\
183 & 140 & 136& 122 \\
223 & 160 & 150& 134\\
\end{tabular}
\end{center}
\end{table}

\clearpage
\newpage

\eject

\section*{FIGURE CAPTIONS}     
\begin{figure}
\begin{center}
\leavevmode
\setlength\epsfxsize{6.0in}
\setlength\epsfysize{7.0in}
\epsfbox{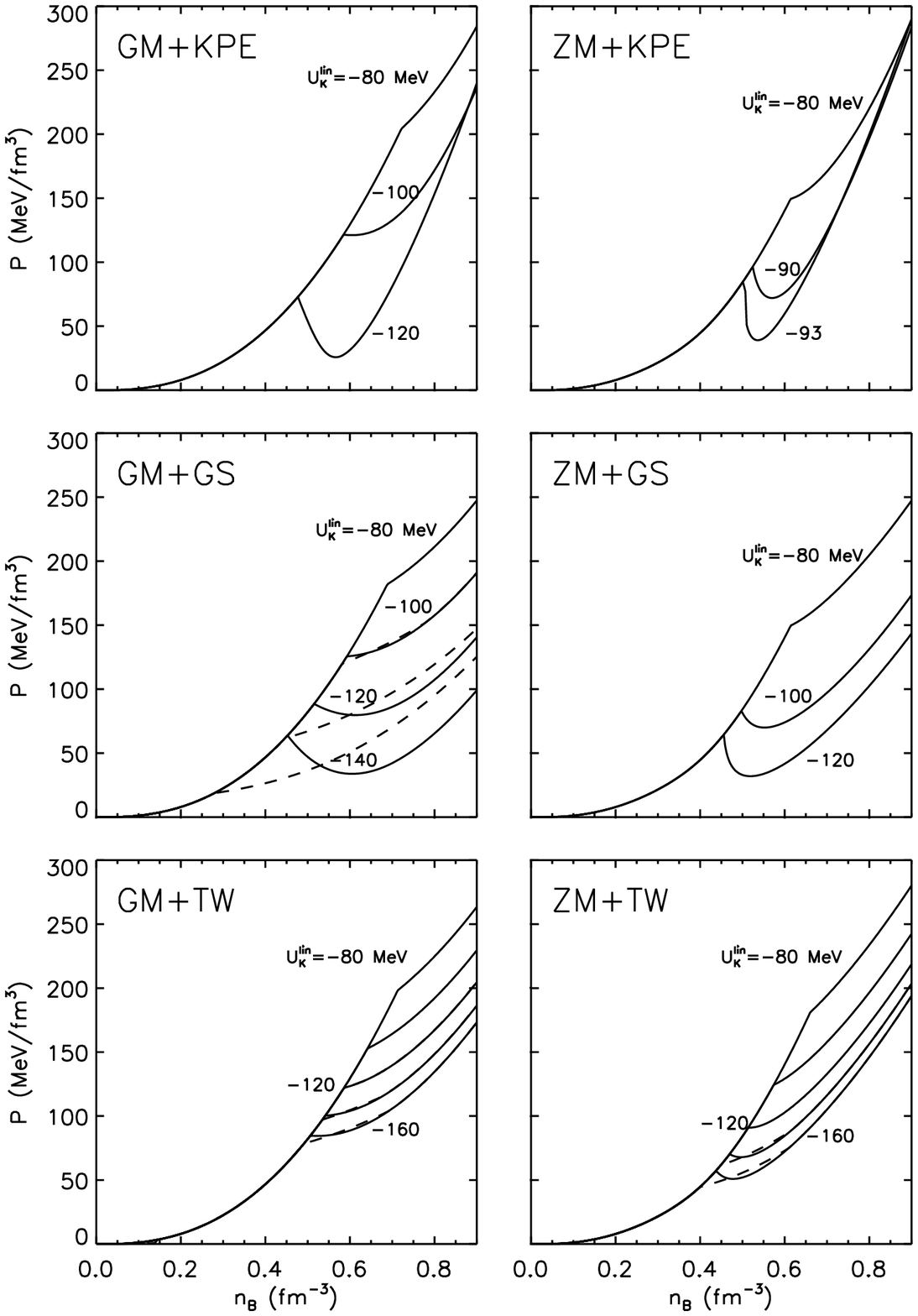}
\caption[]{}
\label{fig1}
\end{center}
\end{figure}

\vskip 10pt FIG. \ref{fig1}: Pressure versus baryon number density for
the six choices of the nucleon and kaon Lagrangians considered in this
paper.  The temperature $T=0$ and there are no trapped neutrinos
($Y_\nu=0$).  Selected values for the kaon optical potential
$U_K^{\rm lin}$ are indicated. The solid lines show the pressure in the
pure phases I (nucleons only) and II (the high-density nucleon-kaon
condensed phase). The dashed lines show the pressures obtained by
imposing Gibbs' criteria for phase equilibrium in a mixed-phase region
for the case of first order transitions.  For the KPE choice of the
kaon Lagrangian, Gibbs' criteria could not be satisfied despite the
occurence of first order phase transitions in some cases.

\begin{figure}
\begin{center}
\leavevmode
\setlength\epsfxsize{6.0in}
\setlength\epsfysize{7.0in}
\epsfbox{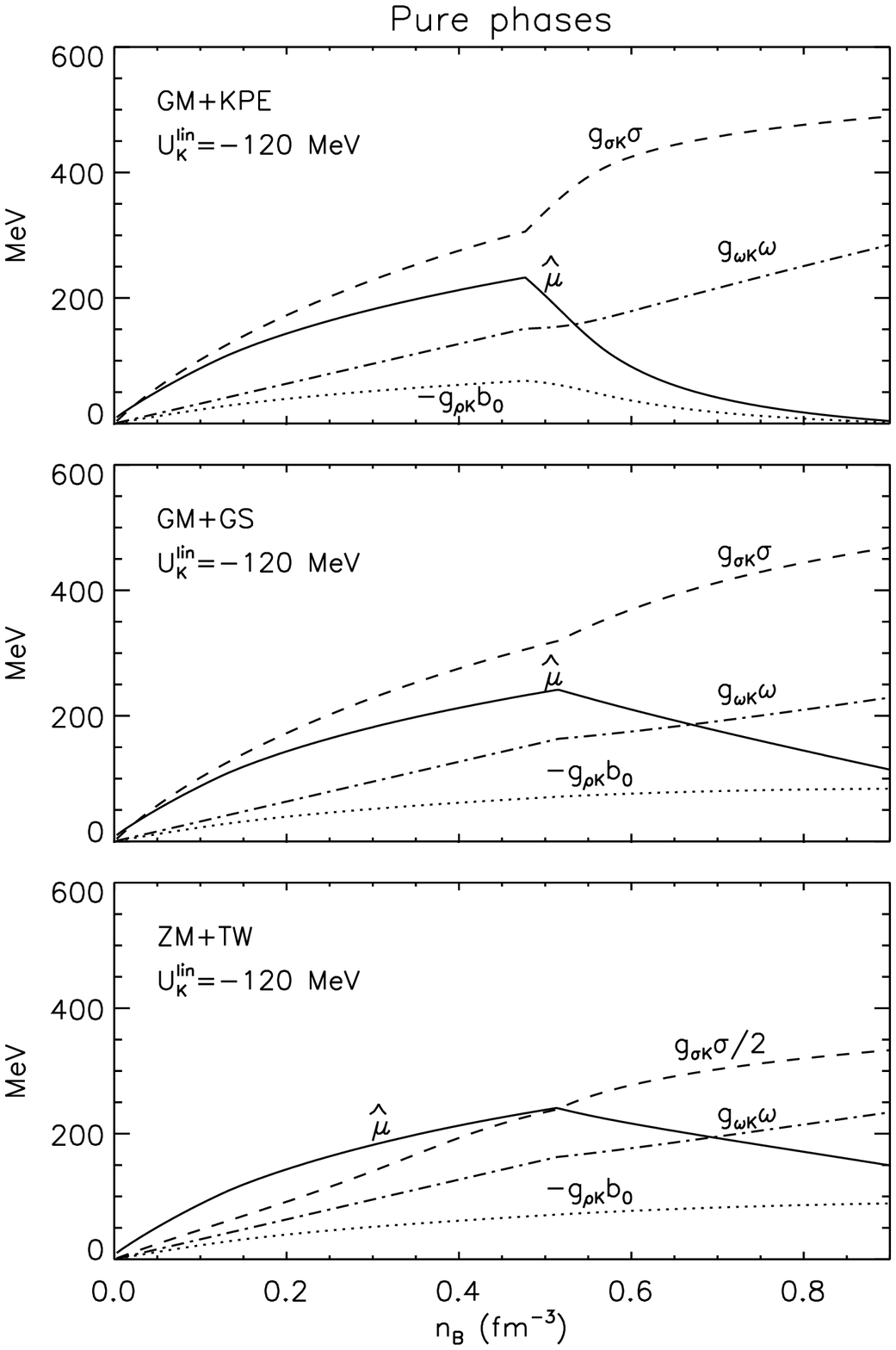}
\caption[]{}
\label{fig2}
\end{center}
\end{figure}

\vskip 10pt
FIG. \ref{fig2}: The density dependences of the scalar, vector, and
iso-vector fields for different choices of the nucleon and kaon Lagrangians
($T=0, Y_\nu=0$).
The solid curves show the chemical potential                              
$\mu_n-\mu_p = \mu$.  In this figure, results are shown only for the pure 
phases I and II; the mixed phase produced by satisfying Gibbs' criteria is
ignored.

\begin{figure}
\begin{center}
\leavevmode
\setlength\epsfxsize{6.0in}
\setlength\epsfysize{7.0in}
\epsfbox{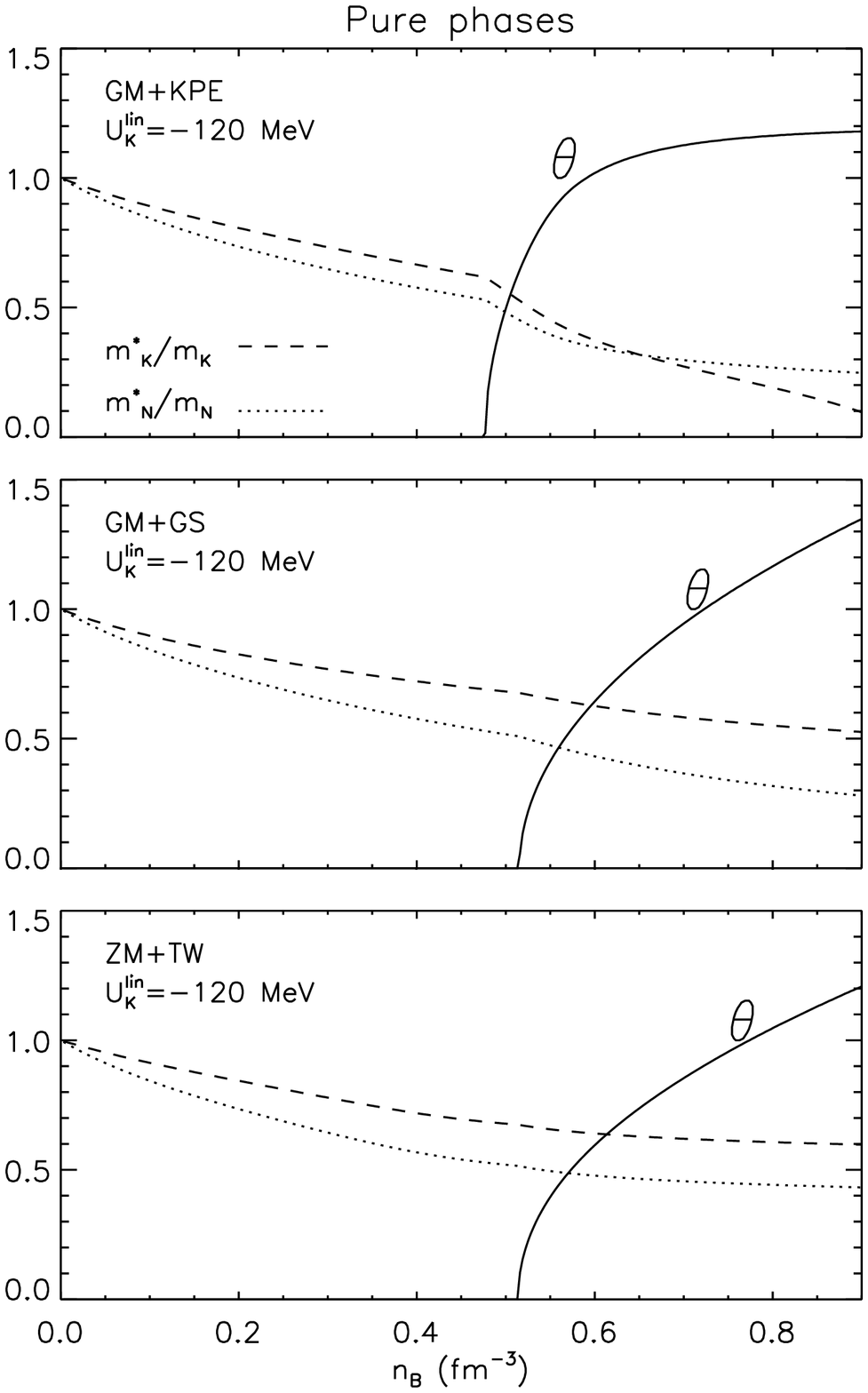}
\caption[]{}
\label{fig3}
\end{center}
\end{figure}

\vskip 10pt FIG. \ref{fig3}: As for Fig. \ref{fig2}, but for the
density dependences of the kaon and nucleon effective masses. The
solid curves show the condensate amplitude.

\begin{figure}
\begin{center}
\leavevmode
\setlength\epsfxsize{6.0in}
\setlength\epsfysize{7.0in}
\epsfbox{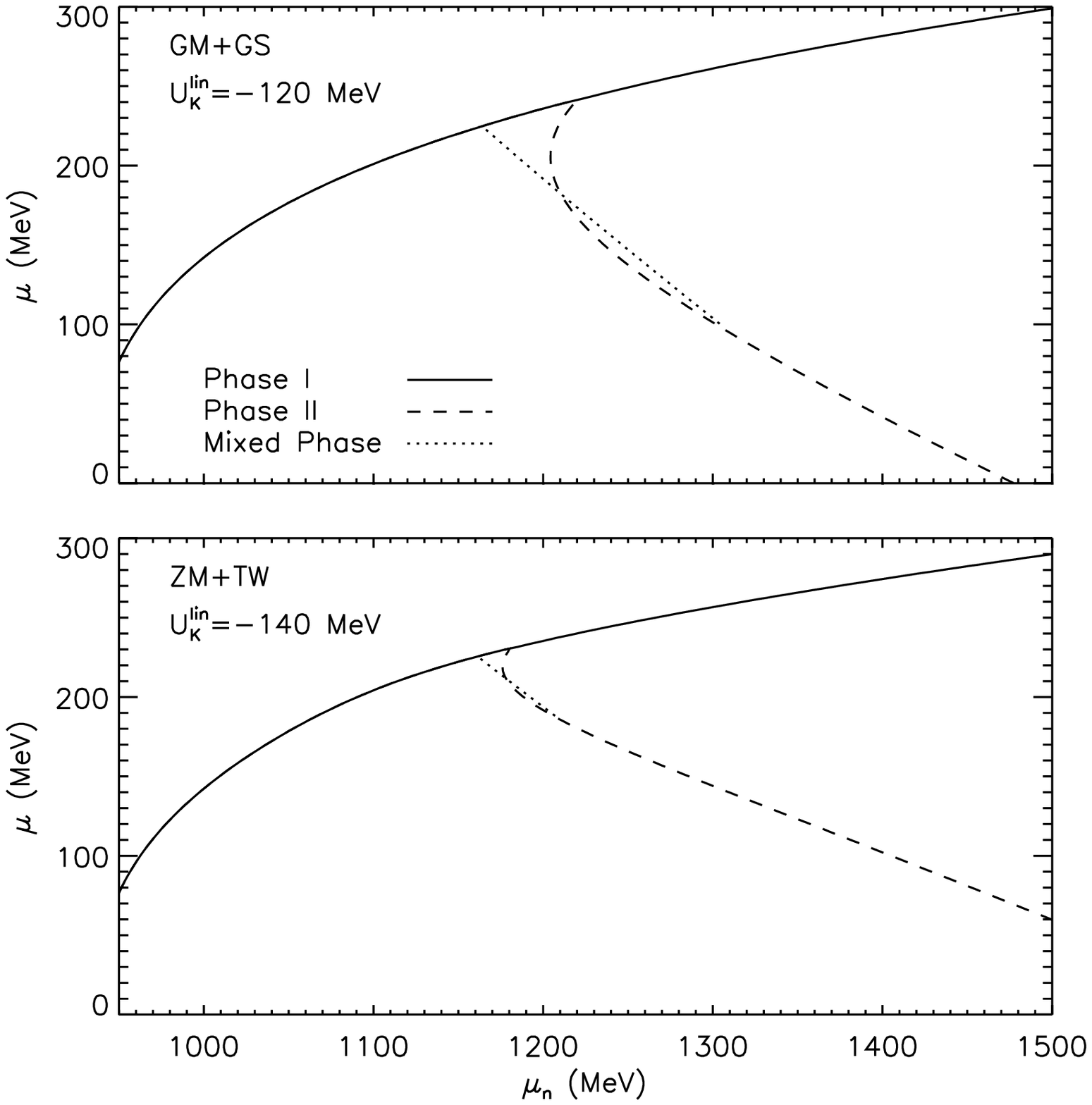}
\caption[]{}
\label{fig3a}
\end{center}
\end{figure}

\vskip 10pt
FIG. \ref{fig3a}:
The electron chemical potential $\mu$ versus  the neutron chemical 
potential $\mu_n$ in pure phases I and II, and in the mixed phase.  
The pure phase I (solid curve) consists of nucleons and leptons. The 
pure phase II (dashed curve) 
is comprised of a kaon condensate coexisting with nucleons and leptons.
The mixed phase (dots) is constructed by satisfying Gibbs' 
rules for phase equilibrium. 

\begin{figure}
\begin{center}
\leavevmode
\setlength\epsfxsize{6.0in}
\setlength\epsfysize{7.0in}
\epsfbox{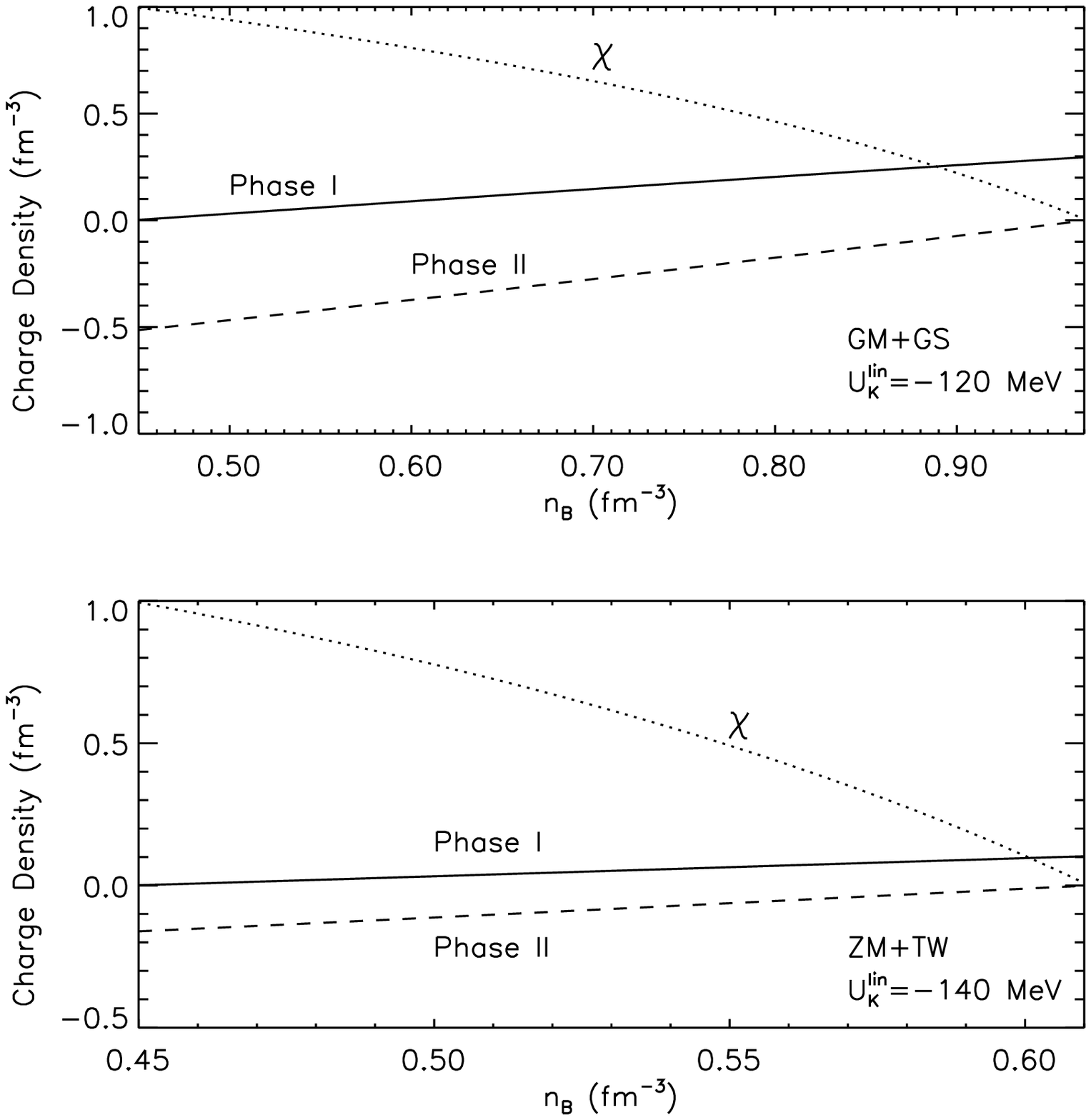}
\caption[]{}
\label{fig3b}
\end{center}
\end{figure}

\vskip 10pt
FIG. \ref{fig3b}: Individual charge densities of pure phases I and II 
and the volume fraction $\chi$ of phase I in the mixed phase 
as a function of baryon density.  Results are for the GM+GS model with 
$U_K^{\rm {lin}} = -120$ MeV and for ZM+TW model with 
$U_K^{\rm {lin}} = -140$ MeV.

\begin{figure}
\begin{center}
\leavevmode
\setlength\epsfxsize{6.0in}
\setlength\epsfysize{7.0in}
\epsfbox{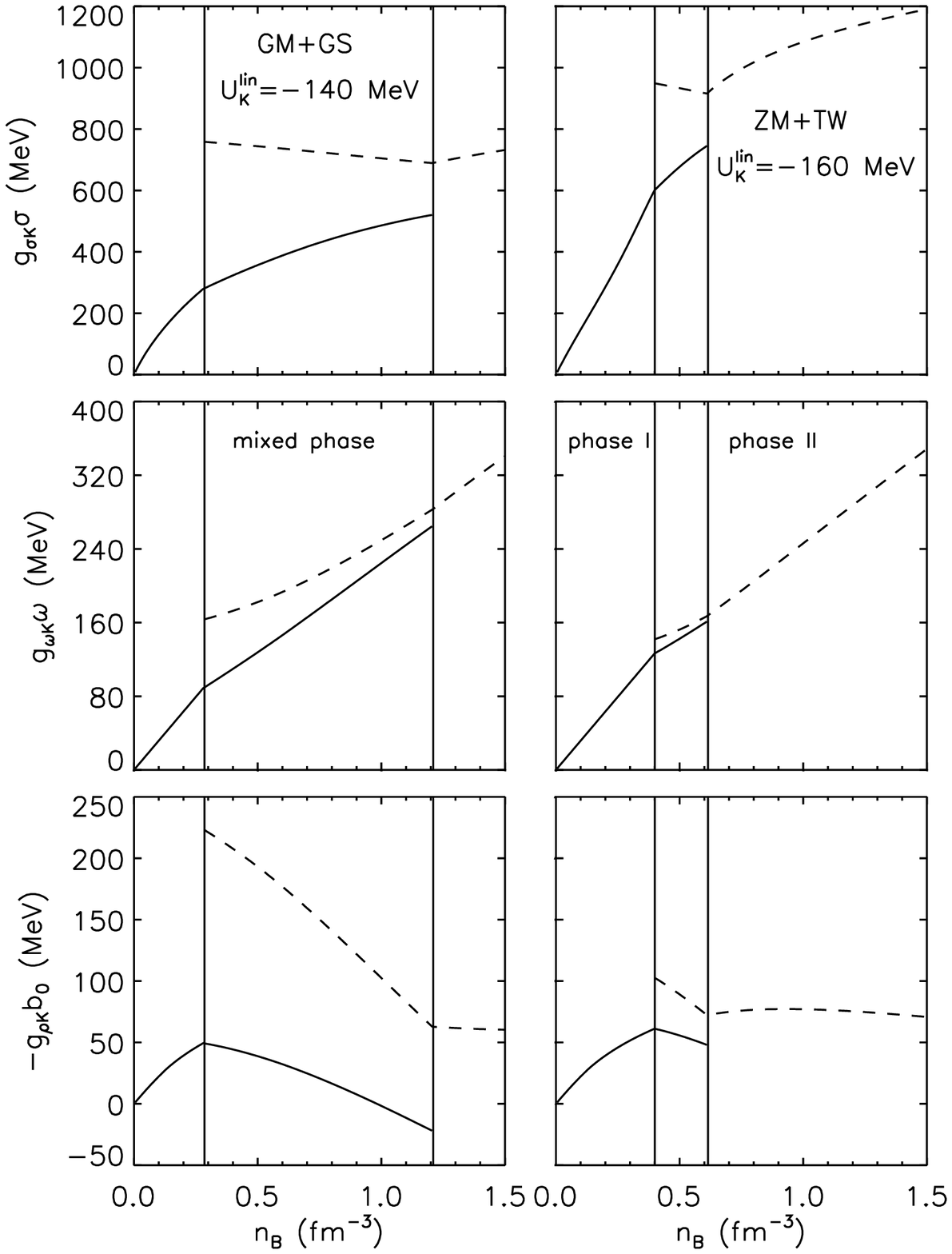}
\caption[]{}
\label{fig4}
\end{center}
\end{figure}

\vskip 10pt
FIG. \ref{fig4}: 
The density dependences of the scalar, vector, and iso-vector fields for
two choices of the nucleon and kaon Lagrangians ($T=0, Y_\nu=0$).  Phase I is 
the pure nucleon phase and phase II is the high-density nucleon-kaon 
condensed phase. The vertical lines demark the mixed phase region.

\begin{figure}
\begin{center}
\leavevmode
\setlength\epsfxsize{6.0in}
\setlength\epsfysize{7.0in}
\epsfbox{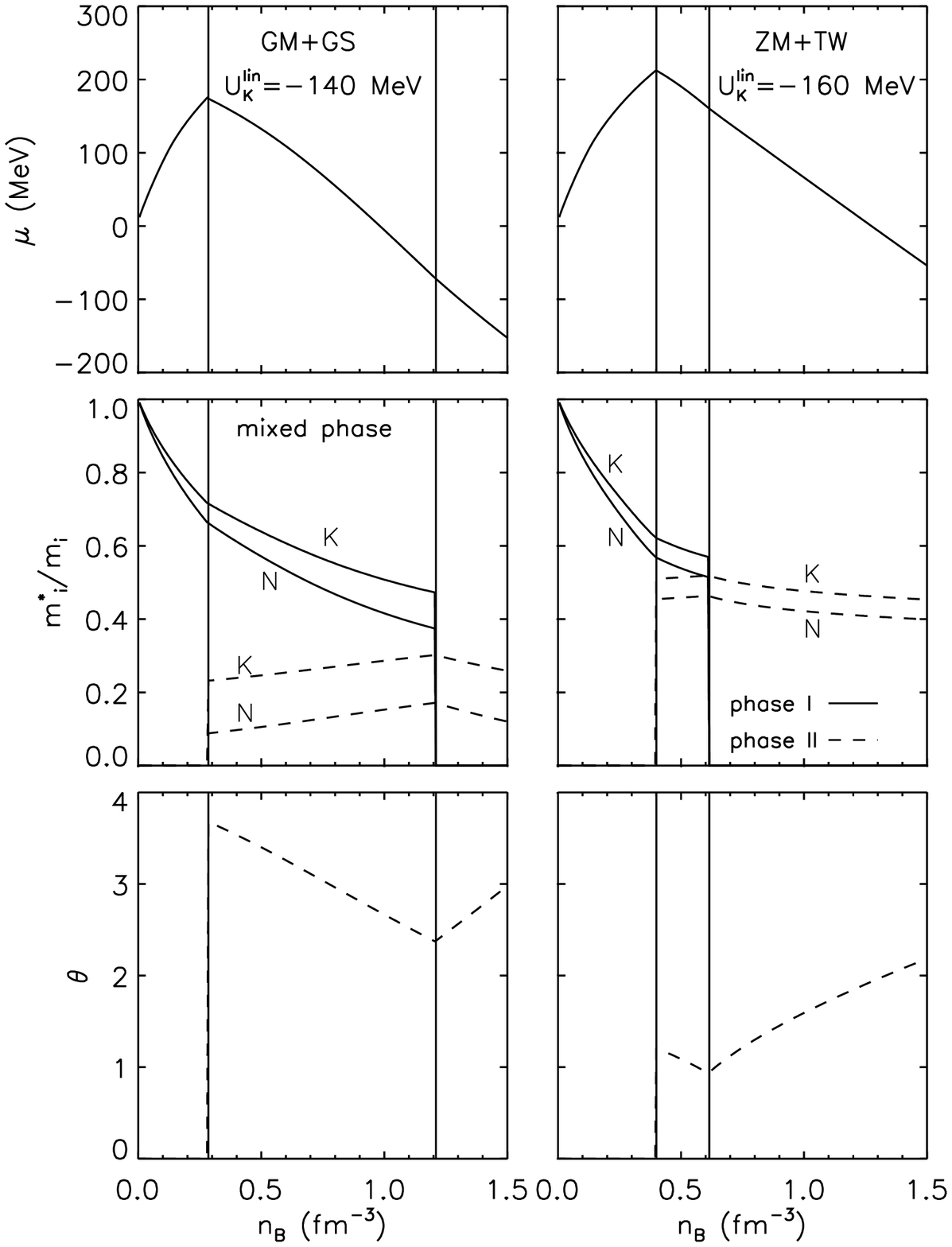}
\caption[]{}
\label{fig5}
\end{center}
\end{figure}

\vskip 10pt
FIG. \ref{fig5}: 
The density dependences of the chemical 
potential $\mu_n-\mu_p=\mu$,  
the kaon (K) and nucleon (N) effective masses, and the condensate amplitude 
($T=0, Y_\nu=0$).  Notation is as in Fig.~\ref{fig4}.

\begin{figure}
\begin{center}
\leavevmode
\setlength\epsfxsize{6.0in}
\setlength\epsfysize{7.0in}
\epsfbox{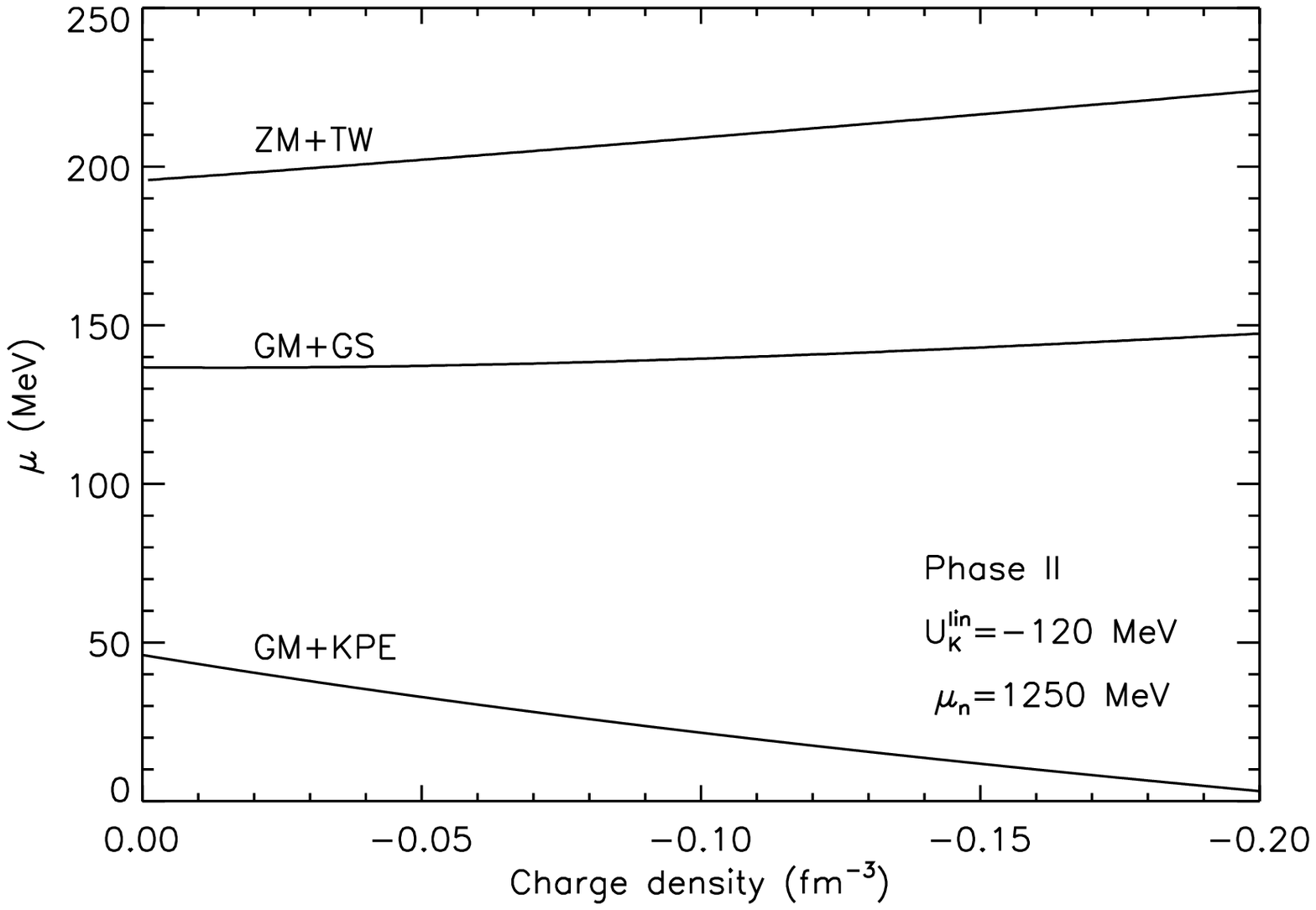}
\caption[]{}
\label{fig5a}
\end{center}
\end{figure}

\vskip 10pt
FIG. \ref{fig5a}:
The electron chemical potential $\mu$ in phase II  
matter versus 
charge density for different models at a fixed neutron chemical 
potential of $\mu_n=1250$ MeV.  In all cases, the optical potential 
$U_K^{\rm {lin}}=-120$ MeV. 

\begin{figure}
\begin{center}
\leavevmode
\setlength\epsfxsize{6.0in}
\setlength\epsfysize{7.0in}
\epsfbox{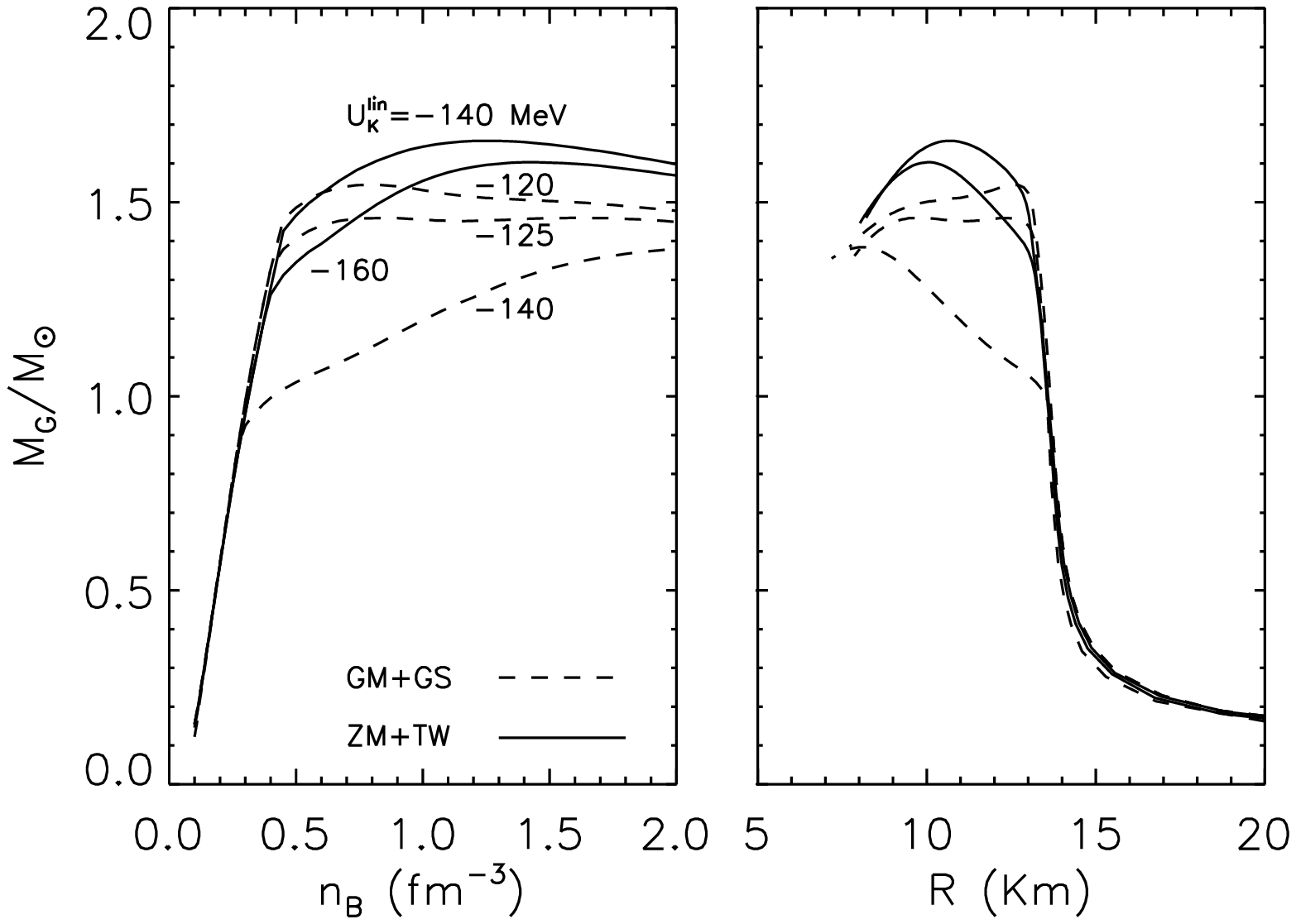}
\caption[]{}
\label{fig6}
\end{center}
\end{figure}

\vskip 10pt
FIG. \ref{fig6}: 
Left panel:  The gravitational mass as a function of the  central
baryon number density for the cases GM+GS and ZM+TW             
($T=0, Y_\nu=0$). Curves are labelled by the values of $U_K^{\rm lin}$
and the EOS includes a mixed phase region.
Right panel: The gravitational mass as a function
of the stellar radius.

\begin{figure}
\begin{center}
\leavevmode
\setlength\epsfxsize{6.0in}
\setlength\epsfysize{7.0in}
\epsfbox{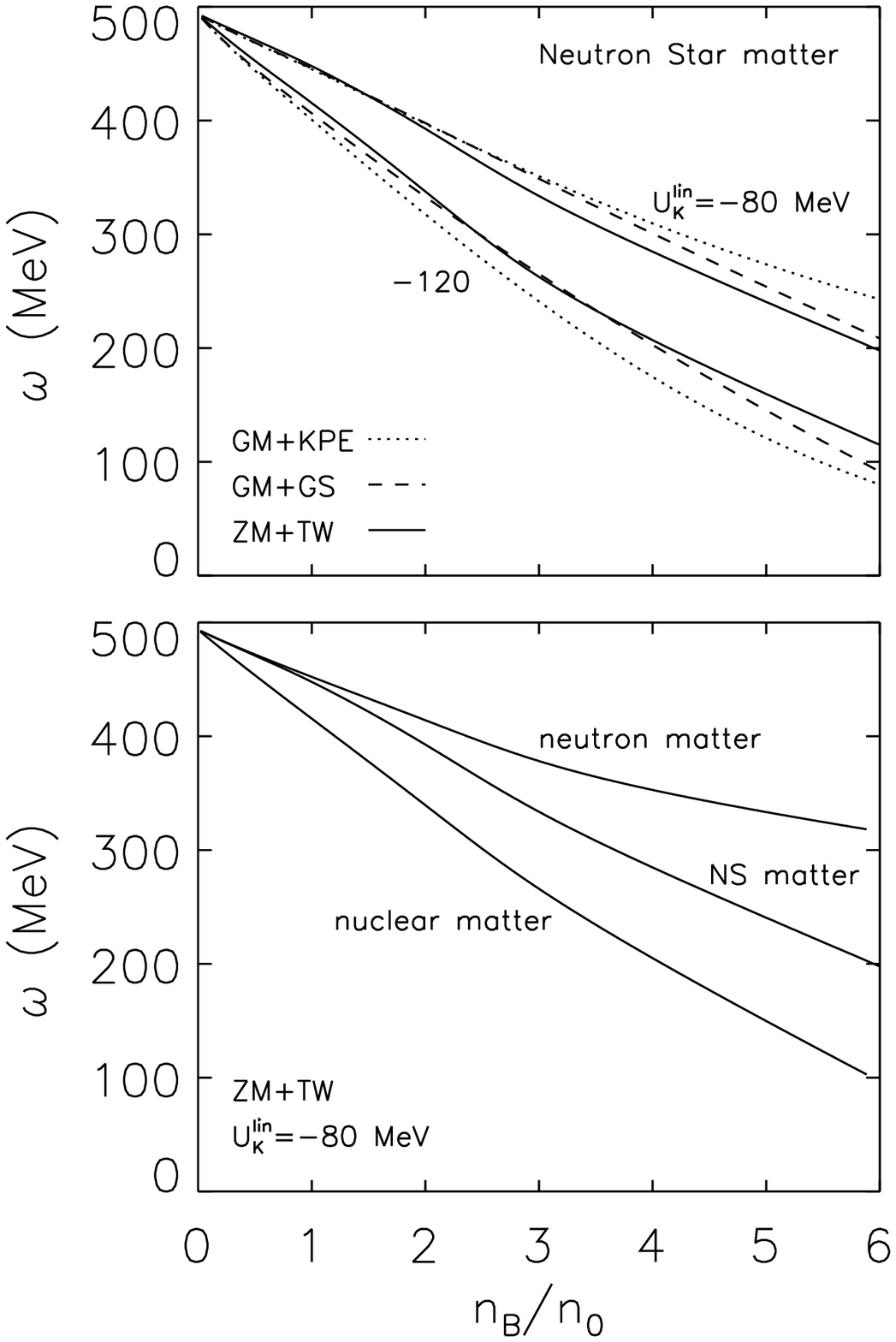}
\caption[]{}
\label{newfig}
\end{center}
\end{figure}

\vskip 10pt
FIG \ref{newfig}: 
The density dependences of the kaon energy $\omega$ in matter with 
different isospin content. The top panel compares results of 
GM+KPE, GM+GS and ZM+TW models for beta
stable neutron star matter for $U_K^{\rm {lin}}=-80$ and -120 MeV,
respectively.  The bottom panel shows results for the ZM+TW model
with  $U_K^{\rm {lin}}=-80$ MeV in pure neutron matter, beta stable
neutron star matter and nuclear matter.

\begin{figure}
\begin{center}
\leavevmode
\setlength\epsfxsize{6.0in}
\setlength\epsfysize{7.0in}
\epsfbox{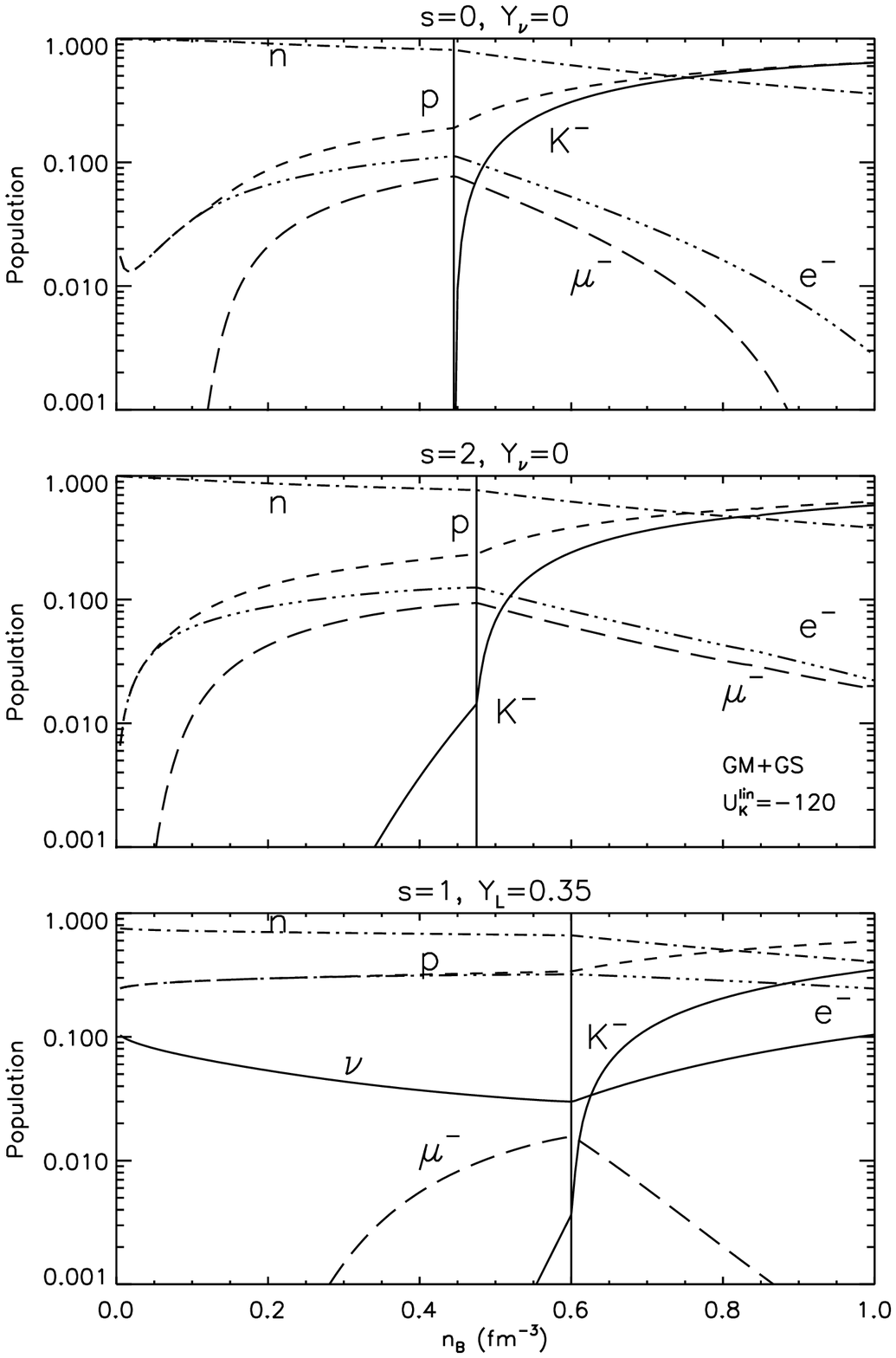}
\caption[]{}
\label{fig7}
\end{center}
\end{figure}

\vskip 10pt
FIG \ref{fig7}: 
The relative concentrations of hadrons and leptons as functions of
baryon number density for three representative snapshots during the evolution
of a PNS.  The results shown are for the model GM+GS. To the left of the 
vertical line there is no kaon condensate, to the right a mixed phase
is present.

\begin{figure}
\begin{center}
\leavevmode
\setlength\epsfxsize{6.0in}
\setlength\epsfysize{7.0in}
\epsfbox{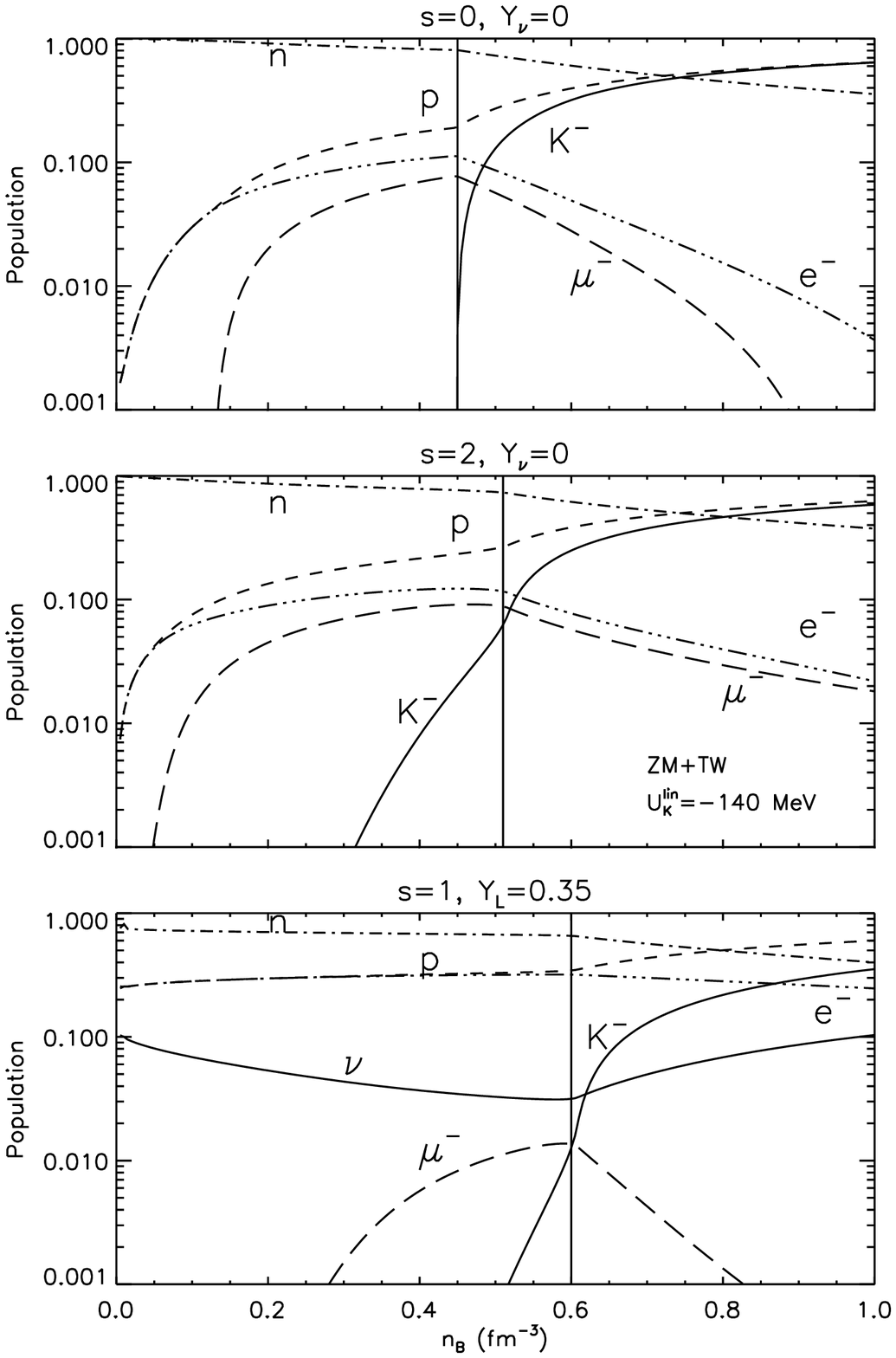}
\caption[]{}
\label{fig8}
\end{center}
\end{figure}

\vskip 10pt
FIG \ref{fig8}: Same as Fig. \ref{fig7}, but for the model ZM+TW.

\begin{figure}
\begin{center}
\leavevmode
\setlength\epsfxsize{6.0in}
\setlength\epsfysize{7.0in}
\epsfbox{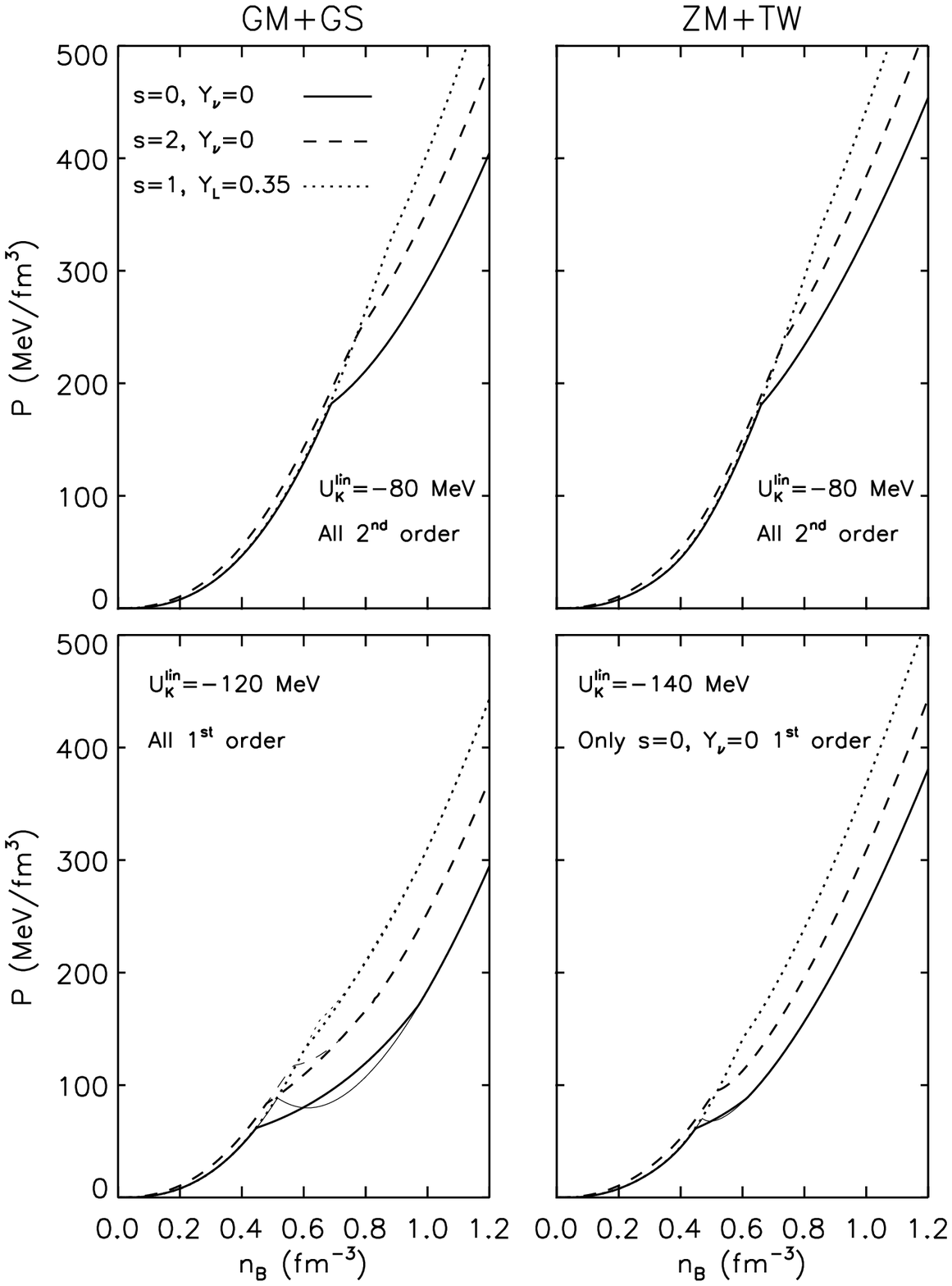}
\caption[]{}
\label{fig9}
\end{center}
\end{figure}

\vskip 10pt FIG \ref{fig9}: The pressure versus baryon number density
for three representative snapshots during the evolution of a PNS.  The
cases shown in the upper panels produce only second order phase
transitions.  For the cases in the lower panels the transitions are
first order, except for ZM+TW with $s>0$.  In the lower panels, heavy
curves include a mixed phase region and light curves ignore a mixed
phase region.

\begin{figure}
\begin{center}
\leavevmode
\setlength\epsfxsize{6.0in}
\setlength\epsfysize{7.0in}
\epsfbox{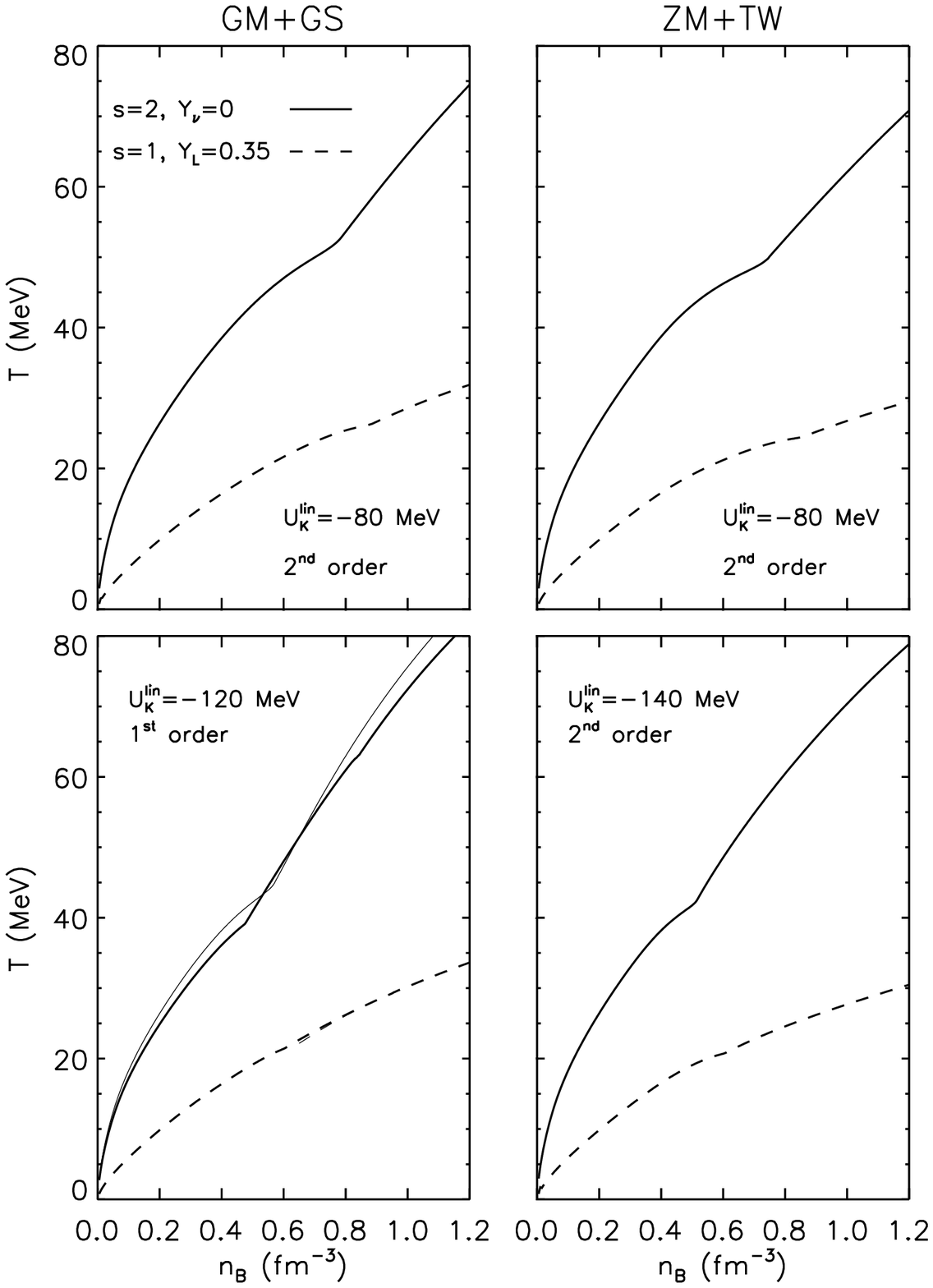}
\caption[]{}
\label{fig10}
\end{center}
\end{figure}

\vskip 10pt FIG \ref{fig10}.  The temperature as a function of baryon
density for two snapshots during the PNS evolution.  Other notation is
as in Fig.~\ref{fig8}.

\begin{figure}
\begin{center}
\leavevmode
\setlength\epsfxsize{6.0in}
\setlength\epsfysize{7.0in}
\epsfbox{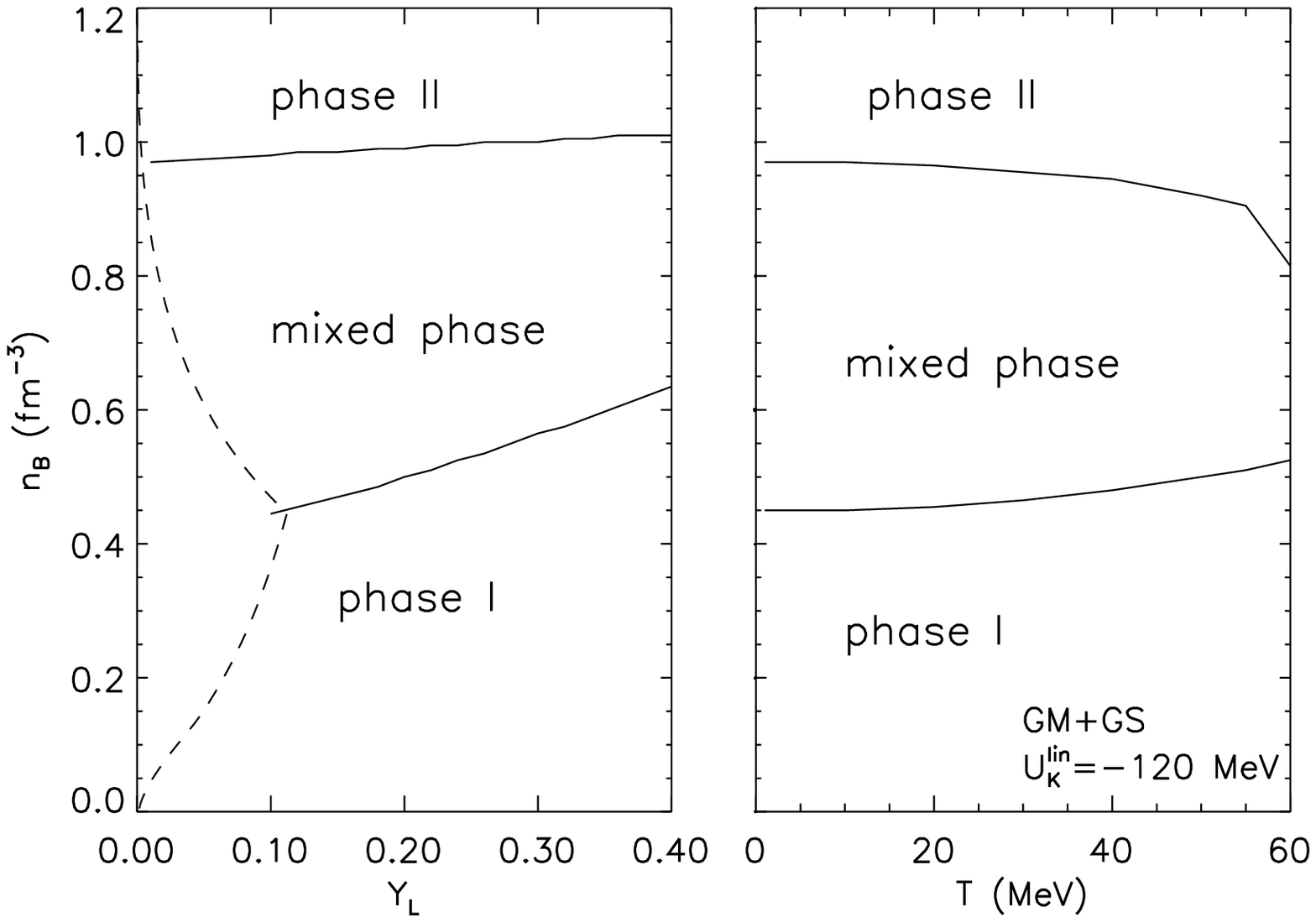}
\caption[]{}
\label{fig11}
\end{center}
\end{figure}

\vskip 10pt
FIG \ref{fig11}: 
The phase diagram of kaon condensed matter for the case GM+GS and
$U_K^{\rm lin}=-120$ MeV. The left panel shows  
results at zero temperature in the density versus lepton concentration plane.
The dashed curve shows the minimum lepton concentration for each density, which
occurs for trapped neutrino concentration $Y_\nu=0$.
The right panel shows results in the density versus temperature plane
for neutrino free matter ($Y_\nu=0$).

\begin{figure}
\begin{center}
\leavevmode
\setlength\epsfxsize{6.0in}
\setlength\epsfysize{7.0in}
\epsfbox{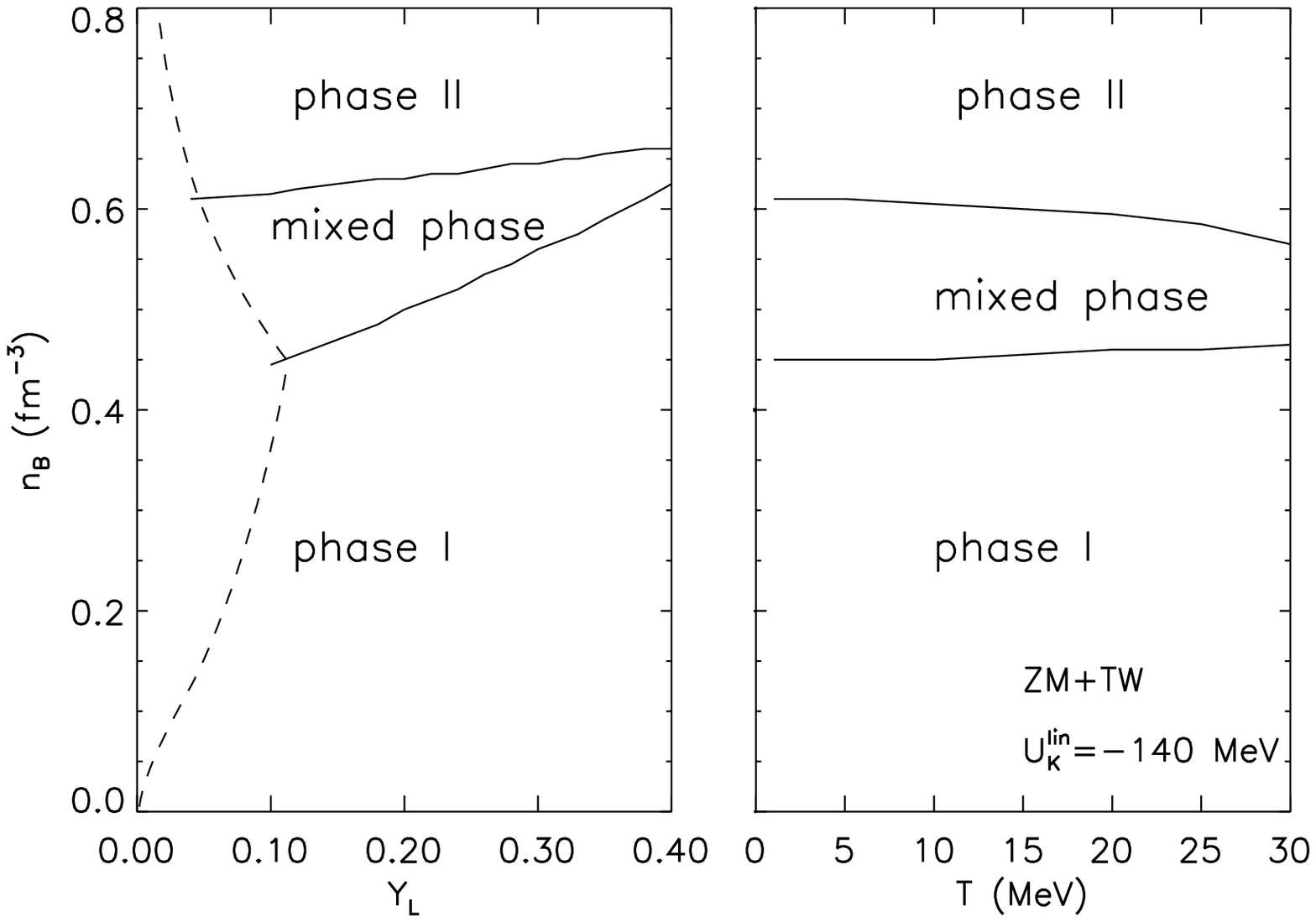}
\caption[]{}
\label{fig12}
\end{center}
\end{figure}

\vskip 10pt
FIG \ref{fig12}: Same as Fig.~\ref{fig11}, but for the model ZM+TW  
and $U_K^{\rm lin}=-140$ MeV. 

\begin{figure}
\begin{center}
\leavevmode
\setlength\epsfxsize{6.0in}
\setlength\epsfysize{7.0in}
\epsfbox{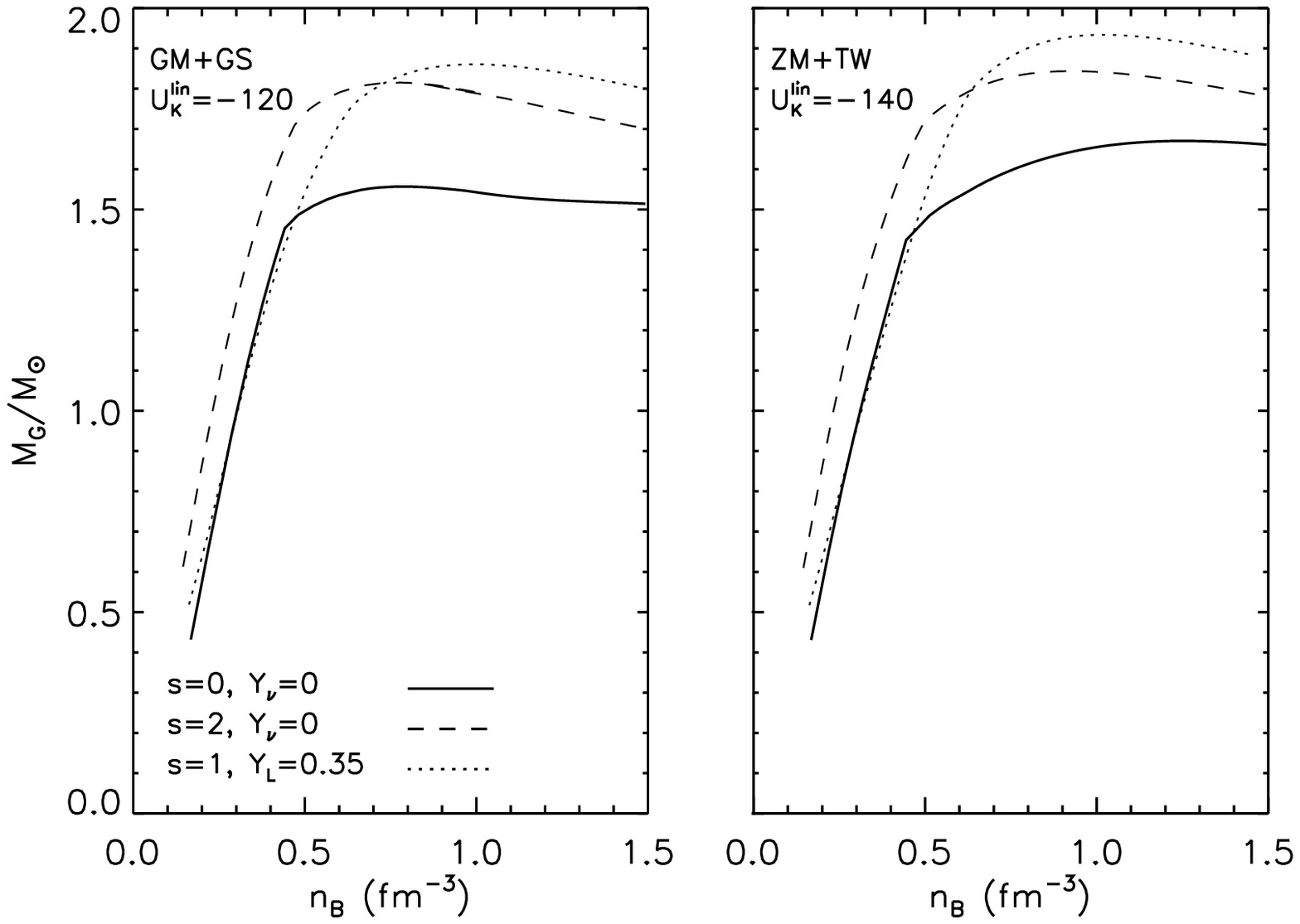}
\caption[]{}
\label{fig13}
\end{center}
\end{figure}

\vskip 10pt FIG \ref{fig13}.  The gravitational mass versus central
baryon number density in the GM+GS and ZM+TW models for three
representative snapshots during the PNS evolution.

\begin{figure}
\begin{center}
\leavevmode
\setlength\epsfxsize{6.0in}
\setlength\epsfysize{7.0in}
\epsfbox{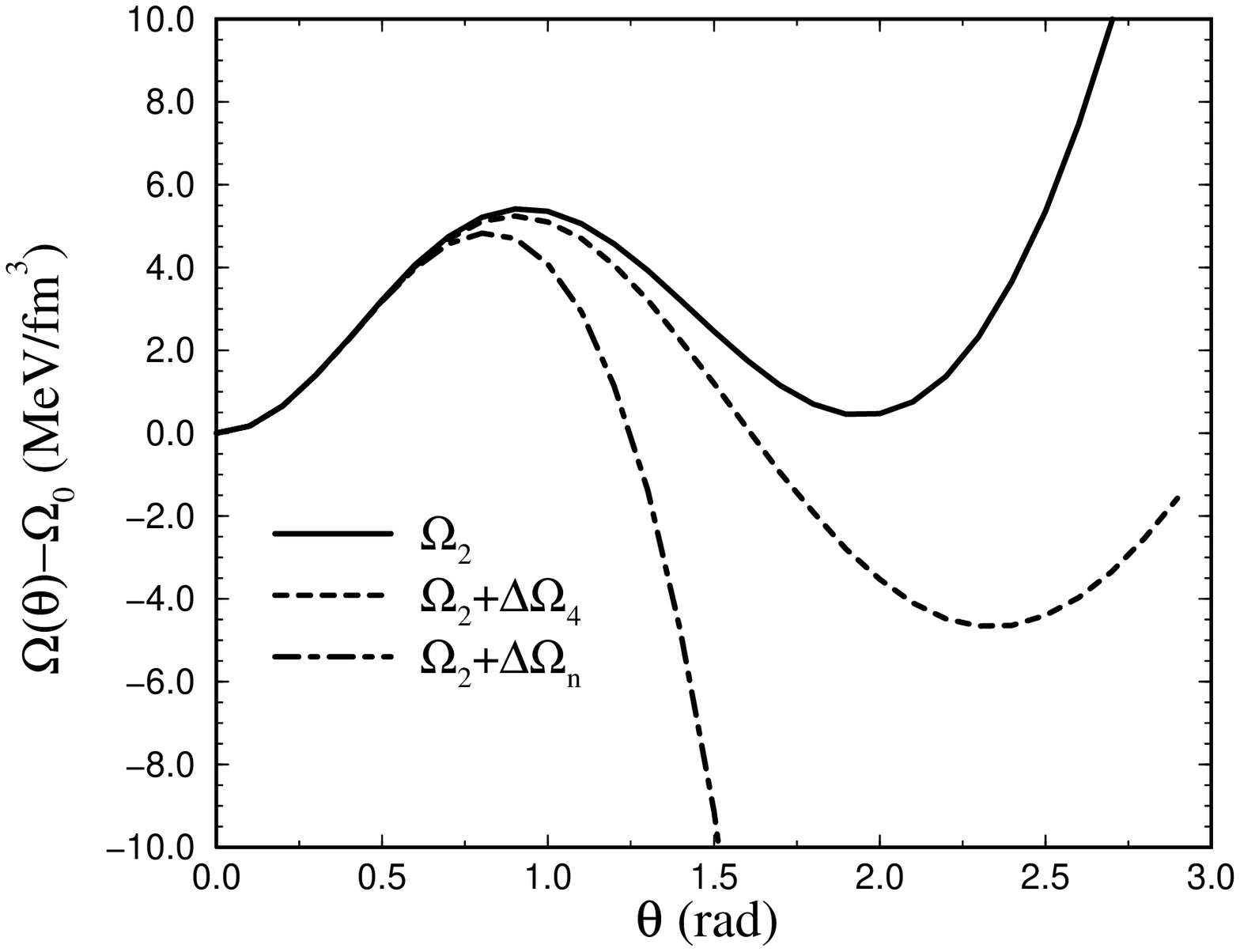}
\caption[]{}
\label{fig14}
\end{center}
\end{figure}

\vskip 10pt
FIG \ref{fig14}. The thermodynamic potential as a function of the 
condensate order parameter $\theta$. 
Results are shown for the GM+GS model near the critical 
density ($\mu_n \simeq 1160$ MeV and $\mu \simeq 243$ MeV) with optical
potential $U_K^{\rm lin}=-120$ MeV. 

\newpage

\newpage

\newpage

\newpage

\newpage

\newpage

\newpage

\newpage

\newpage

\newpage

\newpage

\newpage

\newpage

\newpage

\newpage

\newpage

\newpage

\newpage

\end{document}